\documentclass[
    reprint,              % Format ähnlich wie finale Journal-Version
    amsmath, amssymb,     % Basis-Mathe-Symbole direkt in der Klasse aktivieren
    aps,                  % American Physical Society Format
    prb,                  % Physical Review B Style
    twocolumn,            % Zweispaltiges Layout
    superscriptaddress,   % Autoradressen als Hochstellung
    floatfix,             % Verhindert falsch platzierte Floats
    showpacs,             % Zeigt PACS-Nummern an (veraltet, aber manchmal verlangt)
]{revtex4-2}              % Dokumentklasse für APS-Journale
\usepackage{graphicx}      % Einfügen von Bildern (png, pdf, jpg, eps)
\usepackage{dcolumn}       % Tabellen-Spalten am Dezimalpunkt ausrichten
\usepackage{bm}            % Fette mathematische Symbole
\usepackage{tabularx}      % Tabellen mit variabler Spaltenbreite
\usepackage{multirow}      % Mehrzeilige Zellen in Tabellen
\usepackage{float}         % Verbesserte Steuerung der Float-Positionierung (Option [H])
\usepackage{tikz}          % Zum Zeichnen von Grafiken (z.B. Skizzen, Symbole)
\usepackage{amsmath}       % Erweiterte Matheumgebungen (align, gather, split etc.)
\usepackage{amssymb}       % Zusätzliche Mathe-Symbole
\usepackage{gensymb}       % Gradzeichen (\degree), \ohm usw.
\usepackage{appendix}      % Zusätzliche Kontrolle über Anhänge
\usepackage[T1]{fontenc}   % Saubere Ausgabe von Umlauten, Akzenten
\usepackage[utf8]{inputenc}% UTF-8 Eingabe (bei pdflatex; XeLaTeX/LuaLaTeX brauchen es nicht)
\usepackage[english]{babel}% Sprachunterstützung (Silbentrennung, Bezeichner)
\usepackage{color,soul}    % Farbiger Text, Hervorhebungen (\hl{})
\usepackage{natbib}        % Flexible Zitate, Pflicht für APS/REVTeX
\usepackage{url}           % URLs korrekt umbrechen im Literaturverzeichnis
\usepackage{textcase}      % Korrekte Groß-/Kleinschreibung bei Titeln
\usepackage{hyperref}
\hypersetup{
    colorlinks=true,       % Farbige Links statt Kästen
    linkcolor=blue,        % Farbe interner Links
    citecolor=blue,        % Farbe für Zitate
    filecolor=blue,        % Farbe für Dateilinks
    urlcolor=blue,         % Farbe für URLs
    pdftitle={Growth-Controlled Twinning and Magnetic Anisotropy in CeSb2},
}

\begin{document}

%\preprint{APS/123-QED}

\title{Growth-Controlled Twinning and Magnetic Anisotropy in CeSb$_2$}
% Force line breaks with \\
\author{Jan~T.~Weber}
\email[Corresponding author: ]{jan.weber@stud.uni-frankfurt.de}
\affiliation{Kristall- und Materiallabor, Physikalisches Institut, Goethe-Universität Frankfurt, Max-von-Laue Straße 1, 60438 Frankfurt am Main, Germany}
\affiliation{Ames National Laboratory, U.S. DOE,
Ames, Iowa 50011, USA}
\author{Kristin~Kliemt}
\affiliation{Kristall- und Materiallabor, Physikalisches Institut, Goethe-Universität Frankfurt, Max-von-Laue Straße 1, 60438 Frankfurt am Main, Germany}
\author{Sergey~L.~Bud'ko}
\affiliation{Ames National Laboratory, U.S. DOE,
Ames, Iowa 50011, USA}
\affiliation{Department of Physics and Astronomy, Iowa State University, Ames, Iowa 50011, USA}
\author{Paul~C.~Canfield}
\affiliation{Ames National Laboratory, U.S. DOE,
Ames, Iowa 50011, USA}
\affiliation{Department of Physics and Astronomy, Iowa State University, Ames, Iowa 50011, USA}
\author{Cornelius~Krellner}
\affiliation{Kristall- und Materiallabor, Physikalisches Institut, Goethe-Universität Frankfurt, Max-von-Laue Straße 1, 60438 Frankfurt am Main, Germany}


\begin{abstract}
Cerium diantimonide (CeSb$_2$) is a layered heavy-fermion Kondo lattice material that hosts complex magnetism and pressure-induced superconductivity. The interpretation of its in-plane anisotropy has remained unsettled due to structural twinning, which superimposes orthogonal magnetic responses.  
Here we combine controlled crystal growth with magnetization and rotational magnetometry to disentangle the effects of twinning. Nearly untwinned high-quality single crystals reveal the intrinsic in-plane anisotropy: The in-plane easy axis saturates at $M_{\text{easy}}(4~\text{T}) \approx 1.8~\mu_{\text{B}}$/Ce, while the in-plane hard axis magnetization is strongly suppressed, nearly linear, and comparable to the out-of-plane response. These results resolve long-standing discrepancies in reported magnetic measurements, in which in-plane metamagnetic transition fields and saturation magnetization varied significantly across previous studies.
Growth experiments demonstrate that avoiding the proposed $\alpha$--$\beta$ structural transition---through Sb-rich flux and slower cooling---systematically reduces twinning. However, powder x-ray diffraction and differential thermal analysis measurements show no clear evidence of a distinct $\beta$ phase.
Our results establish a consistent magnetic phase diagram and provide essential constraints for crystal-electric field models, enabling a clearer understanding of the interplay between anisotropic magnetism and unconventional superconductivity in CeSb$_2$.
\end{abstract}

\keywords{CeSb2; Magnetic anisotropy;
Kondo-lattice}
                             
\maketitle

%\tableofcontents

\section{Introduction}

Cerium diantimonide (CeSb$_2$) is a heavy-fermion, Kondo-lattice intermetallic that has attracted significant attention during the last decades due to its anisotropic and rich magnetic phase diagram \cite{Canfield_CeSb2_resis_suszept_1991, Budko_CeSb2_magnetic_1998, Perez_CeSb2_specific_heat_2013, Luccas_CeSb2_heat_capacety_LaSb2_CDW_2015,Zhang_CeSb2_magnetization_2017, Liu_CeSb2_neutron_scattering_2020, Trainer_CeSb2_magnetization_2021}. The recent discovery of pressure-induced unconventional superconductivity beyond the Pauli limit \cite{Squire_CeSb2_pressure_2023} and field-induced magnetic easy-axis switching that persists up to room temperature \cite{Miyake_CeSb2_magnetic_switch_2025} has reinvigorated interest in this material and the broader family of quasi-two-dimensional rare-earth diantimonides, placing CeSb$_2$ again at the forefront of current research into quantum materials with competing magnetic, superconducting, and structural degrees of freedom.

\begin{figure}
     \centering
         \includegraphics[width=\columnwidth]{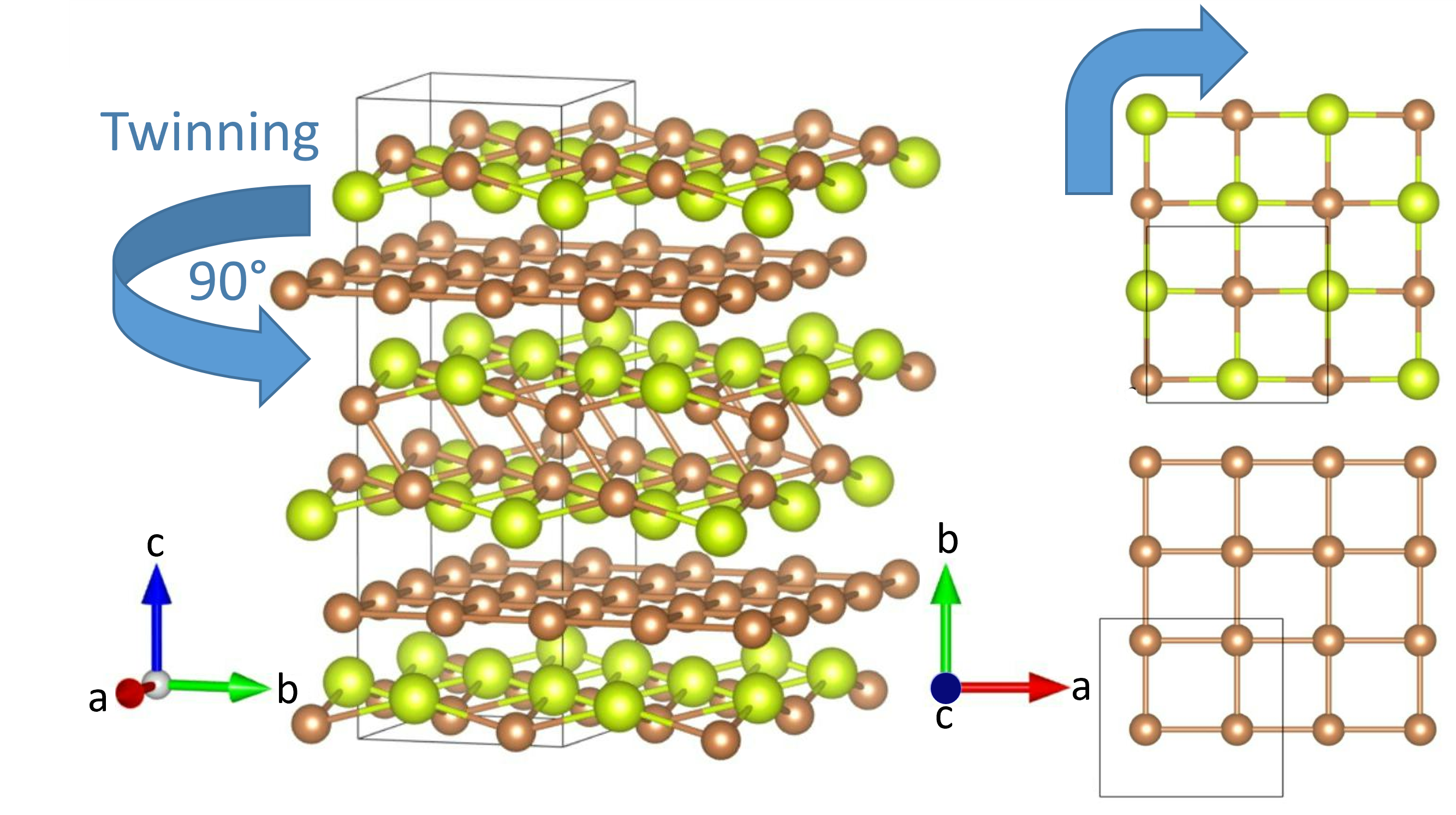}
        \caption{Layered crystal structure of CeSb$_2$ showing Ce--Sb bilayers (green: Ce, brown: Sb) and Sb sheets along the $c$ axis. Twinning is schematically explained (blue). Structure model generated using the \textsc{VESTA} software package \cite{Momma_VESTA_2011}.}
        \label{fig:CeSb2_structure}
\end{figure}

CeSb$_2$ crystallizes---as do many light lanthanoid (LN) and actinoid (AN) diantimonides (SmSb$_2$, LaSb$_2$, NdSb$_2$ \cite{Wang_CeSb2_structure_1967}, GdSb$_2$, TbSb$_2$, PrSb$_2$ \cite{Eatough_YSb2_structure_synthese_1969}, and NpSb$_2$, AmSb$_2$, PuSb$_2$ \cite{Charvillat_NpSb2_AmSb2_PuSb2_structure_1977})---in the orthorhombic SmSb$_2$-type structure (space group \textit{Cmce}, No. 64) \cite{Wang_CeSb2_structure_1967}.
This structure, shown in Fig.~\ref{fig:CeSb2_structure}, consists of pseudo tetragonal alternating LN/AN--Sb bilayers and antimony sheets stacked along the $c$ axis, with each subsequent layer shifted along the $b$ direction. Although the structure appears nearly tetragonal, all known compounds of this type exhibit a small in-plane anisotropy---less than 3~\%---between the $a$ and $b$ lattice parameters. For CeSb$_2$, the lattice constants are $a = 6.295$~\AA, $b = 6.124$~\AA, and $c = 18.21$~\AA \cite{Wang_CeSb2_structure_1967}, indicating an in-plane lattice parameter anisotropy of approximately 2.8~\%.

This pseudo-tetragonal symmetry promotes the formation of structural twins in the $ab$ plane (schematically shown as a $90^\circ$ rotation about the $c$ axis by blue arrows in Fig.~\ref{fig:CeSb2_structure}). Twinning was already confirmed for LaSb$_2$ \cite{Fischer_LaSb2_twinning_2019}, then also suggested for CeSb$_2$ \cite{Liu_CeSb2_neutron_scattering_2020, Shan_FM_ladder_q1D_2025} and confirmed by Miyake \textit{et al.}~\cite{Miyake_CeSb2_magnetic_switch_2025}, who reported $90^\circ$ in-plane twinning with macroscopic domains and domain walls oriented approximately $45^\circ$ to the principal $a$ and $b$ axes. A weak interlayer coupling consistent with the easy cleavability of CeSb$_2$, might allow independent twinning along the $c$ axis. Such structural twinning complicates the interpretation of directional measurements such as the magnetization $M(H)$.

Interestingly, Abulkhaev \textit{et al.} proposed a high-temperature phase of CeSb$_2$ \cite{Abulkhaev_CeSb2_Phasediagram_1997, Okamoto_CeSb_Phasediagram_2001} similarly to a high-temperature phase found for LaSb$_2$ \cite{Murray_LaSb2_high_temperature_phase_1970}. The structural transition from the orthorhombic $\alpha$-CeSb$_2$ phase to the high-temperature $\beta$-CeSb$_2$ phase occurs at $(900 \pm 10)~^\circ$C and intersects the liquidus line at approximately 93~at.\% Sb, as observed by differential thermal analysis (DTA). Micrographs of samples quenched in water from 900$~^\circ$C show the coexistence of two phases. Powder x-ray diffraction (PXRD) of the quenched samples displays extra peaks that cannot be indexed to $\alpha$-CeSb$_2$ alone; these unassigned reflections may originate from the $\beta$-CeSb$_2$ phase, whose exact structure remains unresolved.

All the LN/AN diantimonides exhibit diverse and anisotropic electronic and magnetic ground states \cite{Budko_CeSb2_magnetic_1998, Homma_magnetic_structure_NpSb2_2007, Singha_NdSb2_magnetic_quasi2D_magnetic_structure_2023}, likely rooted in their discussed characteristic layered structure.
The electronic and magnetic properties of CeSb$_2$ are governed by the interplay between localized Ce 4$f$ electrons, hybridization with conduction bands, and strong spin-orbit coupling from Sb. Kondo behavior is evident in resistivity and susceptibility measurements, with a coherence temperature of $T^* \approx 100$~K \cite{Budko_CeSb2_magnetic_1998, Zhang_CeSb2_magnetization_2017}. Recent high-resolution, temperature-dependent ARPES measurements demonstrated the onset of hybridized $f$-bands below $T^*$ and the formation of a coherent Kondo lattice \cite{Zhang_CeSb2_kondo_entanglement_2022}, thereby resolving earlier discrepancies regarding the localization of the 4$f$ states \cite{Joyce_photoelecton_spectra_CeSb2_1992, Arko_CeSb2_Photoemission_1997}. These findings are consistent with a moderately enhanced Sommerfeld coefficient $\gamma \sim 50$~mJ\,mol$^{-1}$\,K$^{-2}$ \cite{Luccas_CeSb2_heat_capacety_LaSb2_CDW_2015}. The comparatively low value of $\gamma$ relative to other heavy-fermion systems likely reflects the temperature-dependent evolution of the hybridization between $f$ and conduction electrons and the resulting reduction of the density of states at the Fermi level \cite{Zhang_CeSb2_kondo_entanglement_2022}.

Magnetization measurements confirm that the easy axis lies in the $ab$ plane \cite{Budko_CeSb2_magnetic_1998}. At ambient pressure, CeSb$_2$ exhibits a complex magnetic phase diagram with at least three magnetic phases below 16~K. These can be tuned via in-plane magnetic fields. However, the reported in-plane metamagnetic transition fields and the resulting magnetic phase diagrams vary significantly by as much as $\sim$1~T \cite{Budko_CeSb2_magnetic_1998,Zhang_CeSb2_magnetization_2017, Liu_CeSb2_neutron_scattering_2020, Trainer_CeSb2_magnetization_2021}. The nature of the magnetic ground state remains under debate in literature: It may be ferromagnetic \cite{Canfield_CeSb2_resis_suszept_1991, Budko_CeSb2_magnetic_1998,Zhang_CeSb2_magnetization_2017,Liu_CeSb2_neutron_scattering_2020}, ferrimagnetic \cite{Trainer_CeSb2_magnetization_2021}, or canted antiferromagnetic \cite{Liu_CeSb2_neutron_scattering_2020}. Neutron diffraction reveals an unidirectional in-plane propagation vector $\mathbf{k} = (\pm1/6, -1, 0)$ below 9.8~K \cite{Liu_CeSb2_neutron_scattering_2020}, suggesting a modulated magnetic structure. Despite the two-dimensional layered crystal structure, complementary inelastic neutron scattering measurements have revealed quasi-one-dimensional spin excitations, well described by a ferromagnetic spin-ladder model \cite{Shan_FM_ladder_q1D_2025}, suggesting that the magnetic dimensionality is lower than the structural layering would imply.

Although detailed information on the crystal electric field (CEF) scheme of CeSb$_2$ is still lacking, it is expected to play a crucial role in shaping the anisotropic magnetic properties. In the orthorhombic environment of CeSb$_2$, the $J = 5/2$ multiplet of Ce$^{3+}$ is split into three Kramers doublets, with the ground-state doublet corresponding to an effective $S=1/2$ system. However, the specific ground-state wavefunction and the associated level splitting have not yet been reported. Reliable knowledge of the saturation magnetization values along the crystallographic axes is essential to constrain possible CEF scenarios and to identify the ground-state doublet. Since twinning obscures the intrinsic anisotropy, a precise determination of the untwinned easy- and hard-axis magnetization is a critical requirement for any meaningful CEF analysis in CeSb$_2$. In addition, anisotropic Curie-Weiss behavior---specifically the signs and magnitudes of the Weiss temperatures and the effective magnetic moments along the principal in-plane directions---provides valuable insight into the underlying exchange interactions and further constrains possible CEF ground-state configurations.

High-field magnetization measurements have revealed a remarkable in-plane easy-axis switching effect \cite{Miyake_CeSb2_magnetic_switch_2025}. The field-applied principal axis becomes the new easy axis, with orthogonal magnetization suppressed, persisting up to room temperature---indicative of a robust memory effect. This behavior is strongly influenced by twinning: In highly twinned as-cast samples, intrinsic anisotropy is masked, whereas high-field exposed untwinned or partially detwinned samples show clear directional dependence (see Supplemental Material of Ref.~\cite{Miyake_CeSb2_magnetic_switch_2025}).

Under applied pressure, CeSb$_2$ undergoes a profound evolution of its magnetic and electronic properties. As commonly observed in Ce-based compounds, pressure enhances the hybridization between localized 4$f$ electrons and conduction electrons, destabilizing the
magnetic moments. This leads to a suppression of magnetic order \cite{Kagayama_CeSb2_pressure_magnetoresistance_2000, Kagayama_CeSb2_pressure_ferromagnetism_collapse_2005} and eventually drives the system through a structural phase transition \cite{Podesta_Poster_pressure_2022,Squire_CeSb2_pressure_2023, Hodgson_PhDThesis_CeSb2_pressure_2023, Squire_PhDThesis_CeSb2_pressure_2024}.
At higher pressures, CeSb$_2$ enters an unconventional superconducting phase with $T_c = 0.22$~K covering a quantum critical point. Its upper critical field $H_{c2}$ exceeds the Pauli limit and shows a characteristic S-shaped temperature dependence, consistent with spin-triplet pairing \cite{Squire_CeSb2_pressure_2023}. Such behavior is consistent with magnetically mediated triplet superconductivity, as ferromagnetic fluctuations offer a plausible pairing mechanism \cite{Shan_FM_ladder_q1D_2025}. Reduced magnetic dimensionality, such as the quasi-one-dimensional spin dynamics seen in CeSb$_2$, is known to enhance such fluctuations and promote unconventional superconductivity. However, a complete understanding of the superconducting ground state and its microscopic origin remains an open question.

%\textbf{In this work}
In this work, we investigate how growth conditions---particularly the passage through the high-temperature structural phase transition---influence the twinning ratio in CeSb$_2$, which has long complicated the interpretation of its anisotropic magnetic properties. We present magnetization data from untwinned, well-oriented single crystals, revealing the intrinsic in-plane magnetic anisotropy and reconstructing the low-temperature phase diagram along the easy axis. These insights are essential for understanding the ambient-pressure magnetic ground state and lay the groundwork for interpreting the pressure-induced superconducting phase in terms of competing magnetic and electronic correlations.

\section{Experimental Details}

\subsection{Crystal growth}

Single crystals of CeSb$_2$ were grown using high-purity cerium and antimony with an excess of Sb as flux. The starting materials were placed in a crucible, sealed in an evacuated fused silica ampoule, and loaded into a box furnace. After the growth process, the residual flux was removed by centrifugation.

Crystals 1 and 2 originated from the same batch and were grown in an alumina fritted Canfield Crucible Set \cite{Canfield_crucible_2016, LPS_ceramics_Canfield_Setup} using a Ce\,:\,Sb molar ratio of 10\,:\,90 with a total mass of 3.5~g. Crystal~3 was grown under otherwise identical conditions, but using a graphite crucible. The mixture was heated to $1000~^\circ$C at a rate of 100\,K/h, held at that temperature for 1 h to homogenize the melt, and then cooled at 3\,K/h down to $800~^\circ$C, where the excess flux was separated by centrifugation. With this growth protocol, the crystals necessarily crossed the proposed high-temperature $\alpha$--$\beta$ structural transition during solidification.

Crystal~4 was grown in an alumina crucible using a lower cerium concentration (Ce\,:\,Sb = 5\,:\,95, total mass 6~g). The mixture was heated to $1000~^\circ$C at 333\,K/h, followed by a slower ramp of 100\,K/h to $1100~^\circ$C. It was then cooled at 2.1\,K/h down to $675~^\circ$C, where the flux was removed by centrifugation. This protocol may have avoided crossing the structural transition, though this cannot be determined with certainty; we therefore classify crystal~4 as a possible case, with respect to crossing the structural transition during the growth.

Crystals 5, 6, and 7 were synthesized in an alumina crucible using the lowest cerium concentration (Ce\,:\,Sb = 3\,:\,97, total mass 4.5~g). The melt was heated to $1100~^\circ$C at a rate of 100\,K/h and held for 4 h to ensure complete homogenization. It was then cooled to $850~^\circ$C at 100\,K/h, followed by slow cooling at 3\,K/h down to $640~^\circ$C, where the flux was separated by centrifugation. These crystals were grown under conditions expected to avoid the proposed high-temperature transition altogether.

All growths produced large, soft, and malleable platelike crystals with an average size of 3\,mm\,$\times$\,3\,mm\,$\times$\,0.3\,mm, which grow in well-defined layers that are easily cleavable but not exfoliable.

\subsection{Structural, chemical, magnetic and electronic characterization}

PXRD was used to confirm the crystal structure of the samples. The diffraction patterns were recorded using a Bruker D8 FOCUS powder diffractometer in Bragg--Brentano geometry with Cu~K$_\alpha$ radiation. The single crystals were ground to powder at room temperature, and the powder was subsequently sieved to ensure a homogeneous grain size distribution for the measurement.
Energy-dispersive x-ray spectroscopy (EDX) measurements were performed using an EDAX (Ametek) detector without the use of external standards to verify the chemical composition of the single crystals.
The orientation of the crystals was determined with a Laue camera using x-ray radiation from a tungsten anode. The corresponding spectra were simulated using \textsc{QLaue} software \cite{QLaue2007} and \textsc{OrientExpress} software \cite{OrientExpress3.4}. The $ab$ plane is parallel to the surface of the platelike crystals. Blurred Laue spots are most likely due to structural twinning.

Due to the similar $a$ and $b$ lattice parameters, the Laue pattern (not shown) appears quasitetragonal, which makes it impossible to unambiguously distinguish the $a$ and $b$ axes. Nevertheless, it is possible to identify the [100]/[010] reflections and distinguish them from [110] or $[1\overline{1}0]$ reflections.
Therefore, we refer to directions parallel to the $a$ or $b$ axis in twinned crystals as m$_1$ when the easy-magnetization axis is dominant, and m$_2$ when the hard-magnetization axis is dominant. The corresponding $45^\circ$-rotated directions are denoted as $d_1$ and $d_2$, where $d_1$ corresponds to a $45^\circ$ clockwise rotation from m$_1$, and $d_2$ to a $45^\circ$ clockwise rotation from m$_2$.

DTA was carried out using a Netzsch STA~409 simultaneous thermal analyzer. Measurements were performed in an open alumina crucible under low pressure argon atmosphere, employing a type-S thermocouple. Heating and cooling cycles without active cooling were used to probe possible structural transitions in CeSb$_2$. However, measurements with Sb-rich flux as well as with CeSb$_2$ single crystals proved challenging, since evaporated Sb condensed on the thermocouple and affected the thermal signal.

A Quantum Design Physical Property Measurement System (PPMS) was used to investigate the magnetic properties via vibrating sample magnetometry (VSM), four-point resistivity, and heat capacity measurements. A Quantum Design Magnetic Property Measurement System 3 (MPMS3) SQUID magnetometer was used to measure the magnetization in dc scan mode as a function of rotation angle.
\section{Results and discussion}

The literature on CeSb$_2$ presents significant discrepancies in reported in-plane magnetization data, particularly in the location of metamagnetic transitions and the degree of magnetic anisotropy \cite{Budko_CeSb2_magnetic_1998, Zhang_CeSb2_magnetization_2017, Liu_CeSb2_neutron_scattering_2020, Trainer_CeSb2_magnetization_2021}. A major source of this variation is the presence of structural twins in the $ab$ plane, which superimpose the responses of two orthogonal magnetic domains. Untwinned or partially detwinned crystals are therefore essential to accurately determine the intrinsic easy- and hard-axis behavior.  

In the following, we systematically compare magnetization data from crystals with different twinning ratios. This allows us to reconstruct the intrinsic in-plane anisotropy of CeSb$_2$ and to extract a realistic low-temperature magnetic phase diagram along the true easy axis. 
In parallel, we explore different crystal-growth protocols and present PXRD studies, including high-temperature quenching experiments, in an effort to reduce twinning and investigate the possible $\beta$-CeSb$_2$ phase. These complementary results help to interpret the magnetic data in the context of crystal quality and structural variations.

\subsection{Twinning and magnetic anisotropy}
\begin{figure}
\centering
\includegraphics[width=\columnwidth]{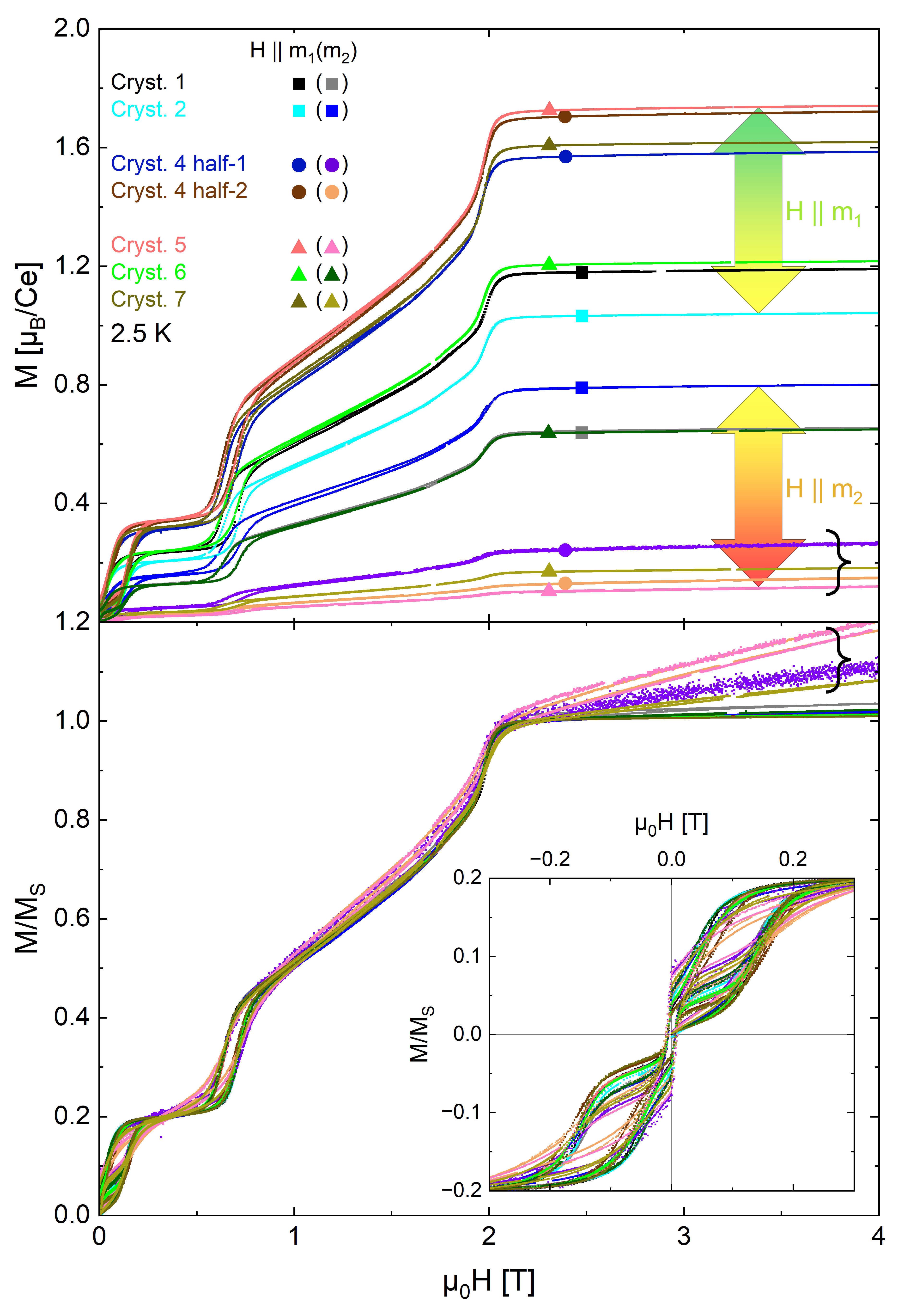}
\caption{Top: In-plane $M(H)$ at 2.5~K for various crystals, showing m$_1$/m$_2$ behavior due to twinning. Symbols denote the different growth conditions: Squares---crystals that crossed the proposed high-temperature transition; circles---crystals that may have crossed it; triangles---crystals that avoided it. Bottom: Same data normalized to the saturation magnetization $M_S=M(2.2~\text{T})$; m$_1$ curves collapse, while low m$_2$ curves (indicated by the bracket) show an additional linear term. Inset: $M(H)$ around small fields; a tiny hysteresis is visible at $\mu_0H=0~\text{T}$ for all crystals.}
\label{fig:M_H_all_curves_and_normalized}
\end{figure}

The effect of twinning on in-plane magnetic anisotropy is illustrated in Fig.~\ref{fig:M_H_all_curves_and_normalized} (top graph), where magnetization isotherms at 2.5~K are shown for several crystals. Crystals 1 and 2 and 5, 6 and 7 were measured as-cast while crystal~4 was cleaved parallel to the $ab$ plane and both halves were measured separately. The measurements were performed with an external magnetic field H along m$_1$ and m$_2$ (with m$_1$ and m$_2$ defined above as the in-plane directions dominated by the easy- and hard-axis domains, respectively). All curves show metamagnetic transitions at $0.10~\text{T}$, $0.66~\text{T}$, and $1.97~\text{T}$. For each crystal, one direction falls into the m$_1$ regime (easy axis in the majority, higher magnetization curve) and the other into the m$_2$ regime (hard axis in the majority, weaker magnetization curve). The superposition of easy- and hard-axis contributions in twinned crystals masks the much weaker intrinsic hard-axis behavior. Thus, it can only be derived from crystals with nearly no twinning.

Normalizing the curves to their saturation magnetization $M_S \approx M(2.2~\text{T})$ reveals that m$_1$ curves from different crystals collapse onto a single curve, consistent with negligible hard-axis contributions (Fig.~\ref{fig:M_H_all_curves_and_normalized}, bottom). In contrast, m$_2$ curves---particularly those with very low magnetization signals, indicative of less twinning (i.e., higher crystal quality)---show an additional linear component and therefore deviate systematically from the other curves. As shown in Fig.~\ref{fig:M_H_all_curves_and_normalized} (bottom, bracket), the relative strength of this linear term increases for crystals with less twinning, demonstrating that it originates from the intrinsic hard-axis response. The same normalization factors applied to $M/M_S$ at 2.5~K also align the temperature-dependent susceptibility $M/H(T)$ at $\mu_0H=0.1$~T in the range $10 \leq T \leq 300$~K up to the paramagnetic (PM) regime, confirming that the easy- and hard-axis anisotropy holds above $T_N \approx 15~\text{K}$ (Appendix~\ref{Appendix:Magnetic susceptibility}, shown up to 30~K).

To derive information about the twinning from the different curves and quantitatively compare them, we define the twinning ratio $x = \frac{\text{fraction of easy-axis domains along } m_1}{\text{total domain population along } m_1}$, with $x \in [0.5,1]$, which quantifies the fraction of the easy-axis domain oriented along the m$_1$ direction. A higher twinning ratio therefore corresponds to reduced twinning and better crystal quality, while a lower ratio indicates stronger twinning and poorer crystal quality.\\
Assuming that the magnetization of the hard axis is negligible compared to that of the easy axis, we estimated the twinning ratio $x$ of a crystal from the saturation magnetization at 2.2~T:
\begin{align}
    x \approx \frac{M_{m_1}(2.2~\text{T})}{M_{m_1}(2.2~\text{T})+M_{m_2}(2.2~\text{T})}.
    \label{eq:x_ratio_approx}
\end{align}
We chose $M(2.2~\text{T})$ as the reference point because it lies well above the saturation field yet remains low enough to minimize contributions from the linear hard-axis component. This provides a consistent and reliable benchmark for comparing different samples. Since the hard-axis contribution is small but finite, this approach yields only a lower bound for the actual twinning ratio with an accuracy of 2~\% (Appendix~\ref{Appendix_Error analysis of lower boundary of x}).

To further validate our $x$ extraction procedure from $M(2.2~\text{T})$, we compared $M_{m_1}(2.2~\text{T})+M_{m_2}(2.2~\text{T})$ across different crystals. Since this sum should equal $M_{\text{easy}}(2.2~\text{T})+M_{\text{hard}}(2.2~\text{T})$ independent of the twinning ratio, any deviations directly reflect calibration or alignment errors. In our measurements, the values agreed within $\sim$4~\%, confirming both the accuracy of our crystal positioning and the robustness of the model. All values for this calculations are provided in Appendix~\ref{Appendix:Measurement_accuracy}.

\begin{figure}
\centering
\includegraphics[width=\columnwidth]{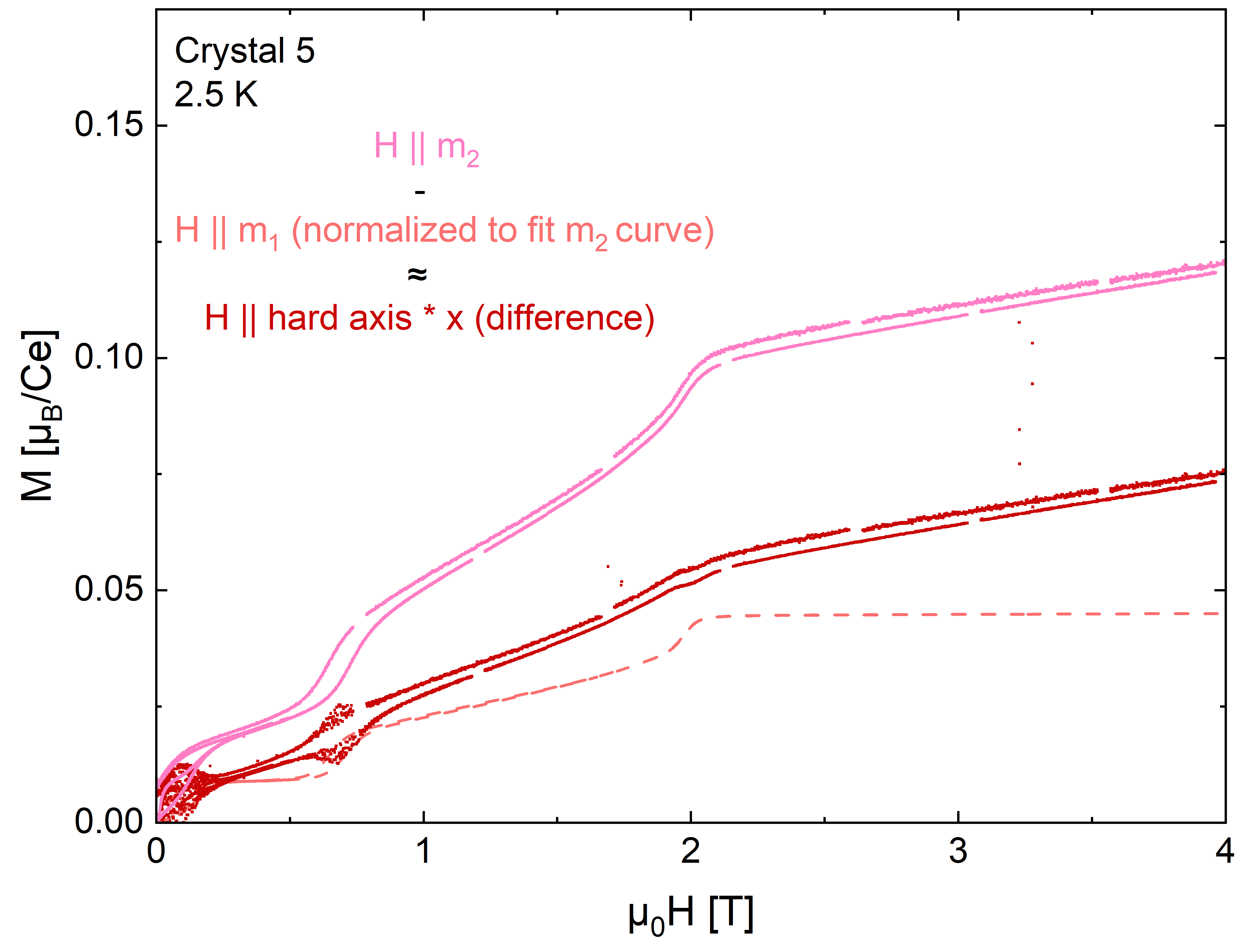}
\caption{Extraction of the intrinsic hard-axis magnetization for crystal~5
by scaling and subtracting the m$_1$ curve from the m$_2$ curve.
A faint hysteresis is visible in the hard-axis response.}
\label{fig:Crystal5_hard_axis_extraction}
\end{figure}

\begin{figure}
\centering
\includegraphics[width=\columnwidth]{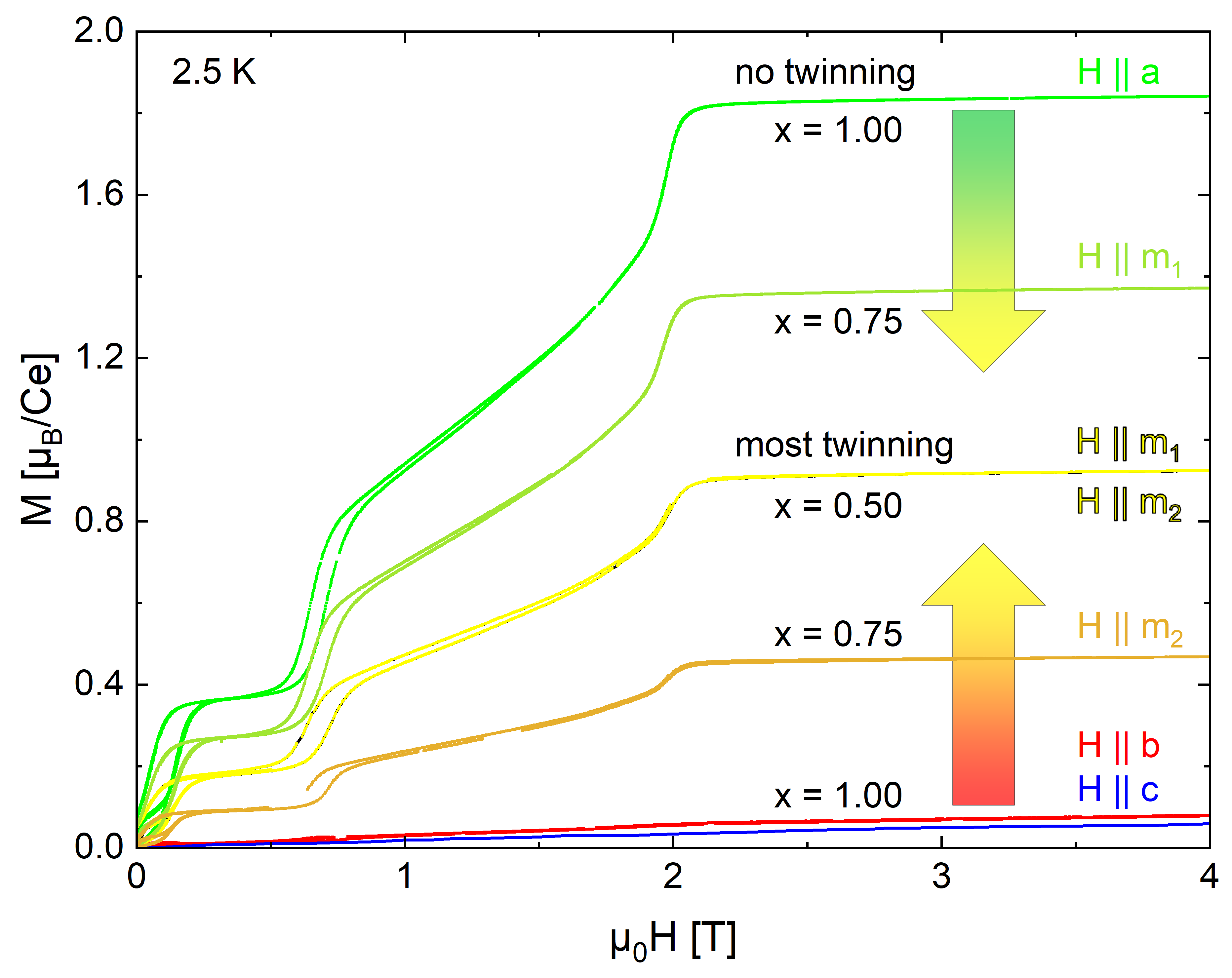}
\caption{Schematic of in-plane magnetization at $T=2.5$~K for varying twinning ratios, including the untwinned anisotropy ($x=1$). The $a$ axis is assigned as the easy axis and the $b$ axis as the hard axis (similar to the $c$-axis response reported by Bud'ko \textit{et al.}~\cite{Budko_CeSb2_magnetic_1998}) \cite{Miyake_CeSb2_magnetic_switch_2025}, though the opposite assignment cannot be excluded \cite{Shan_FM_ladder_q1D_2025}.}
\label{fig:CeSb2_twinning_schematic_real_in_plane}
\end{figure}

To reconstruct the intrinsic hard-axis response, we used crystals with minimal twinning and thus minimal masking of the hard axis. We scaled their m$_1$ curves down by a factor and subtracted it from the corresponding m$_2$ curves. The scaling factor is chosen in such a way that it minimizes features at the m$_1$ transition fields in the difference curve, consistent with an ideally flat linear hard-axis response. This procedure is motivated by several experimental observations. First, metamagnetic features in m$_2$ weaken continuously as the twinning ratio improves and become increasingly dominated by a linear background, indicating that they originate from residual easy-axis contributions rather than from an intrinsic hard-axis response. Second, no shifted metamagnetic transition fields are observed along m$_2$, as would be expected if such transitions were intrinsic to the hard axis in a magnetically anisotropic system. Third, the out-of-plane hard axis ($H \parallel c$) reported by Bud'ko \textit{et al.}~\cite{Budko_CeSb2_magnetic_1998} exhibits a similarly featureless magnetization curve up to 4~T. The resulting flat difference curve is then equivalent to the hard-axis magnetization times the twinning ratio $x$ of the crystal (formalized in Appendix~\ref{Appendix:hard-magnetization}). The method is illustrated in Fig.~\ref{fig:Crystal5_hard_axis_extraction} for crystal~5. 
A small hysteresis, extending approximately from 0.5 to 4~T, which disappears at higher temperatures, is observed in the extracted hard-axis magnetization. Importantly, this hysteresis is already present in the raw m$_2$ data prior to any subtraction procedure, indicating that it is not an artifact introduced by the subtraction method. While its origin cannot be unambiguously confirmed, an experimental contribution cannot be entirely ruled out.
Assuming the qualitative behavior of the hard-axis curve was determined correctly by the difference curve, it still needs to be divided by the corresponding twinning ratio. Given the calculated $x$ is a lower boundary the resulting curve is an upper boundary. The resulting curve for the the intrinsic hard-magnetization axis of CeSb$_2$ is depicted in Fig.~\ref{fig:CeSb2_twinning_schematic_real_in_plane} in red.

To summarize the influence of twinning on measurements along the m$_1$ and m$_2$ directions and to visualize the unmasked intrinsic anisotropy, Fig.~\ref{fig:CeSb2_twinning_schematic_real_in_plane} presents a schematic representation illustrating how twinning modifies the measured curves compared to the true in-plane anisotropy of an untwinned crystal. The curves were reconstructed from experimental data with twinning ratios close to those depicted by scaling the magnetization to the desired twinning ratio. For an untwinned sample ($x=1$), the in-plane easy-axis magnetization reaches 
$1.72\,\mu_{\text{B}}\text{/Ce}\lesssim M_{\text{easy}}(2.2~\text{T})\lesssim 1.82\,\mu_{\text{B}}$/Ce, while the in-plane hard axis is strongly suppressed, with 
$0.04\,\mu_{\text{B}}\text{/Ce}\lesssim M_{\text{hard}}(2.2~\text{T})\lesssim 0.06\,\mu_{\text{B}}$/Ce, and shows no transition features. The definitions of the given lower and upper bounds are provided in Appendix~\ref{Appendix:upper_lower_bound}. The observed hard-axis behavior is comparable to the out-of-plane response ($H\parallel c$) reported by Bud'ko \textit{et al.}~\cite{Budko_CeSb2_magnetic_1998}. While Miyake \textit{et al.}~\cite{Miyake_CeSb2_magnetic_switch_2025} proposed that the $a$ axis corresponds to the in-plane easy axis and the $b$ axis to the hard axis, this assignment remains under debate \cite{Shan_FM_ladder_q1D_2025}.

Having established the upper and lower bounds of the saturation magnetization that define the in-plane magnetic anisotropy relevant for CEF modeling, we now turn to the high-temperature PM regime to probe the underlying exchange interactions. Curie-Weiss fits of the inverse magnetic susceptibility between 125 and 300~K (Appendix~\ref{Appendix:Inverse magnetic susceptibility}) yield similar effective magnetic moments for all measured crystals along m$_1$ and m$_2$, consistent with Ce$^{3+}$ ($4f^1$), but reveal opposite signs of the Curie-Weiss temperatures for the two principal in-plane directions, in agreement with the findings of Miyake \textit{et al.}~\cite{Miyake_CeSb2_magnetic_switch_2025}. Specifically, $\Theta_{\text{CW}}=55$~K for m$_1$ (easy-axis-dominated) and $\Theta_{\text{CW}}\approx -100$~K for m$_2$ (hard-axis-dominated). The anisotropic exchange behavior complements the strong Ising character and provides an important constraint for modeling the CEF ground state.

\subsection{Rotational dependence, model validation, and twinning domains}
To interpret the angular dependence of the magnetization, verify the internal consistency of our data, and reconcile apparent deviations from earlier off-axis measurements reported in the literature, it is essential to construct a quantitative model that accounts for twinning effects in CeSb$_2$.

\begin{figure}
\centering
\includegraphics[width=\columnwidth]{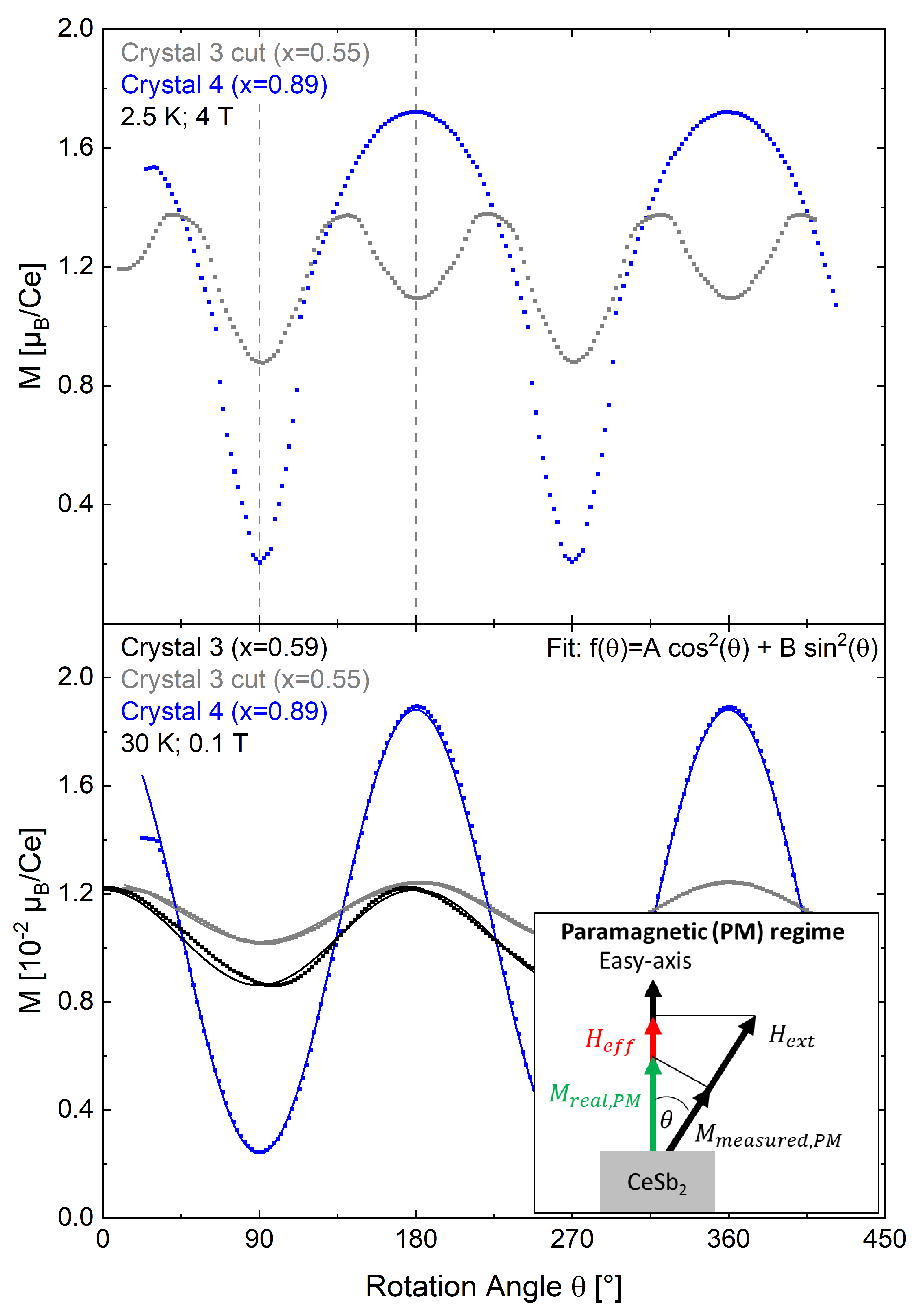}
\caption{Top: Angular dependence of the magnetization $M(\theta)$ measured at 2.5~K and $\mu_0H=4$~T.
Bottom: $M(\theta)$ in the PM regime with corresponding model fits (solid lines). 
Inset: Schematic illustrations of the expected angular dependence for the PM regime.}
\label{fig:rotational_measurements}
\end{figure}
We therefore employ a simple Ising-type model (assuming negligible hard-axis magnetization) with two orthogonal easy axes, weighted by two variables $A$ and $B$. In the field-polarized (FP) regime, where the external field is sufficiently strong to reach $H_\text{S}$ for a restricted range of rotation angles $\theta$, the measured magnetization follows a $|\cos\theta|$ dependence for each domain, resulting in a combined response of $A|\cos\theta|+B|\sin\theta|$ (Appendix~\ref{Appendix:Rotational Dependence Model}). In the PM regime, the angular dependence can be described quantitatively over the full rotation range. The response follows $(\cos\theta)^2$ for a single domain, yielding $A(\cos\theta)^2+B(\sin\theta)^2$ for a twinned crystal (Fig.~\ref{fig:rotational_measurements}, bottom, inset PM model and Appendix~\ref{Appendix:Rotational Dependence Model}). The corresponding twinning ratio can then be expressed as
\begin{align}
    x=\frac{\text{MAX}[A,B]}{A+B}.
\end{align}
Rotational magnetization measurements $M(\theta)$ performed on both highly twinned and nearly untwinned crystals at 2.5~K and an external field of 4~T (Fig.~\ref{fig:rotational_measurements}, top) and at 30~K external field of 0.1~T (Fig.~\ref{fig:rotational_measurements}, bottom) confirm that the proposed Ising-type model accurately reproduces the angular dependence of the data after correcting for small phase and period offsets. For the 2.5~K measurement, the field is sufficient to reach the FP state only within the limited angular windows $29.5^\circ < \theta < 60.5^\circ$, with a periodicity of $90^\circ$ (Appendix~\ref{Appendix:Rotational Dependence Model}), while at other angles the system re-enters the magnetically ordered phase, leading to discontinuities in the behavior of $M(\theta)$. Consequently, we do not attempt to model the complex angular dependence at 2.5~K. Instead, these measurements are used to extract magnetization values along the principal in-plane directions, $M(180^\circ)=M_{m_1}$ and $M(90^\circ)=M_{m_2}$, where the applied field is aligned with the dominant easy and hard axes, respectively. The twinning ratio can then be determined in direct analogy to the analysis of the $M(H)$ curves [Eq.~\eqref{eq:x_ratio_approx}]. The 30~K measurement takes fully place in the PM regime and fits yield twinning ratios perfectly consistent with those extracted from the $M(H)$ saturation values---for instance, crystal~3 gave $x=0.59$ (PM) and $x=0.59$ from $M(H)$. For crystal~4, the comparison of the independently measured 2.5~K rotation measurement and the FP model yielded identical results as well ($x=0.89$ (2.5~K) and $x=0.89$ (PM)), demonstrating the internal consistency of the model across magnetic regimes.

\begin{figure}
\centering
\includegraphics[width=\columnwidth]{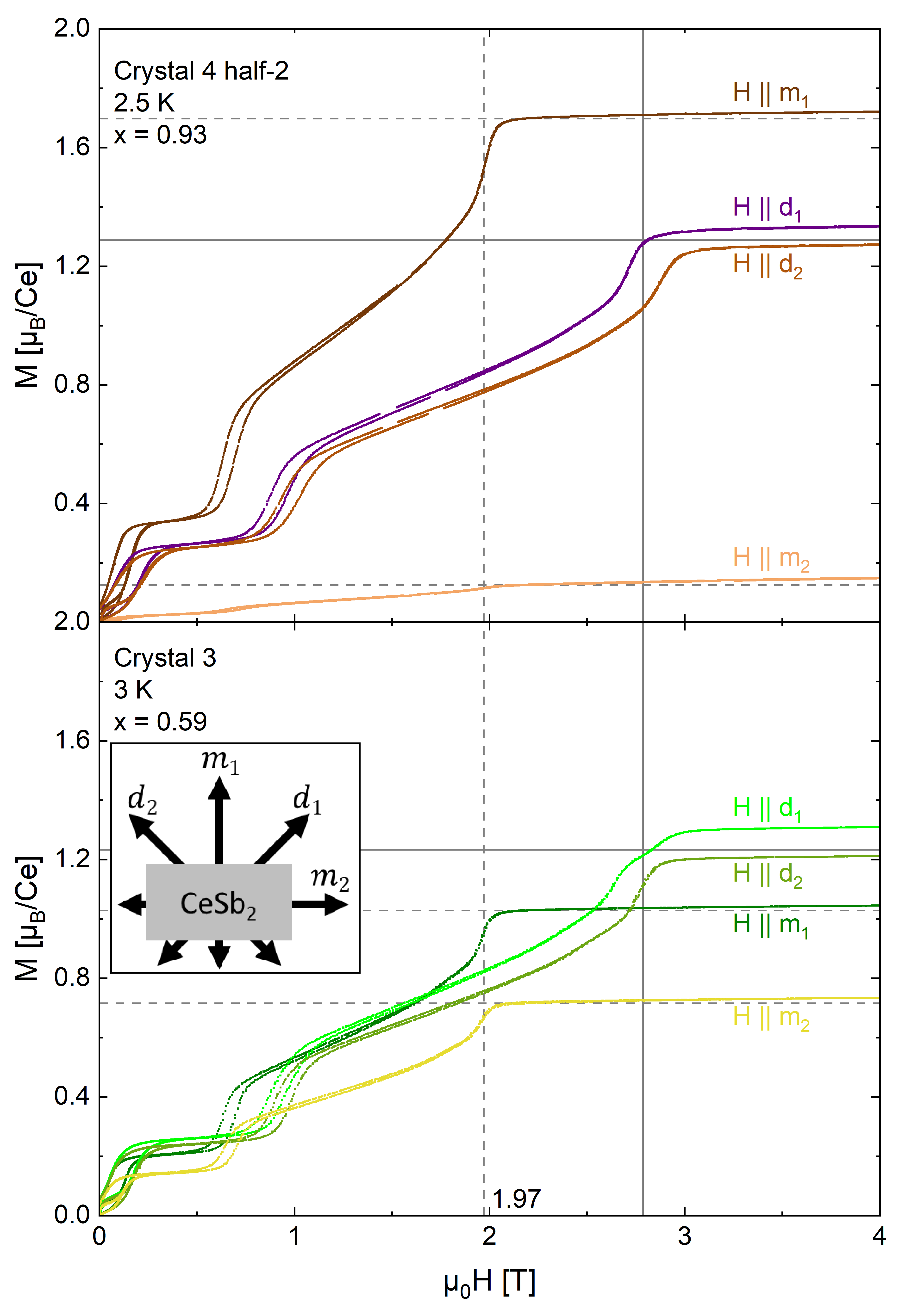}
\caption{Measured $M(H)$ along m$_1$/m$_2$ and d$_1$/d$_2$ for crystals with high and low twinning ratios. $M_S$ and $\mu_0H_S$ for the m$_1$/m$_2$ directions are indicated by dashed gray lines, while the solid gray lines show the corresponding model predictions for d$_1$/d$_2$. Inset: Schematic for m$_1$/m$_2$ and d$_1$/d$_2$ directions.}
\label{fig:Cryst3_4_half_2_d1d2_curves}
\end{figure}

Measurements along directions rotated by $45^\circ$ (d$_1$ and d$_2$) also agree well with the FP-model predictions (solid horizontal lines) when the saturation values $M(2.2~\text{T})$ measured along m$_1$ and m$_2$ (dashed horizontal lines) are used as inputs for the FP model, as shown in Fig.~\ref{fig:Cryst3_4_half_2_d1d2_curves}.

To gain insight into the structure of the twinning domains, we compared twinning ratios of differently cut samples. Crystal~4 ($\approx300~\mu$m thick) was cleaved parallel to the $ab$ plane into two halves, which showed significantly different twinning ratios determined from $M(H)$ measurements: half-1, $x=0.87$; half-2, $x=0.93$. These values bracket the twinning ratios derived for the intact crystal from the rotational measurements ($x=0.89$) as expected. This result, combined with the overall low degree of twinning in the sample, supports the picture of macroscopic, platelike twinning domains---or, at least, independently twinned adjacent layers within the $ab$ plane.

A cut sample from crystal~3 (named crystal~3~cut), representing roughly 10~\% of the original plate surface and which was cut perpendicular to the $ab$ plane, yielded $x=0.55$ in both rotational measurement regions (Fig.~\ref{fig:rotational_measurements}) compared to $x=0.59$ (PM) for the intact crystal. If the twinning domains were perfectly platelike, extending parallel to the $ab$ plane through the full crystal with alternating $90^\circ$ rotations, no change in $x$ would be expected upon cutting perpendicular to the $ab$ plane. The slight variation between the cut and intact values does not exclude this scenario but may also indicate partial twinning within individual plates, as suggested by Miyake \textit{et al.}~\cite{Miyake_CeSb2_magnetic_switch_2025}. Some deviations in $x$ between crystal 3 cut and crystal~3 may also arise from systematic uncertainties in the dc scan mode of the MPMS3, where imperfect centering and deviations from the ideal dipole response function can lead to small scaling errors in the extracted magnetic moment.

To explain not only the differences in saturation magnetization reported in the literature but also variations in transition fields, we next examine the angular dependence of these transitions. The transition fields are expected to follow a simple cosine or sine dependence due to the varying projection of the magnetic field $H_\text{eff}$ along the easy axis. Such angular shifts are already visible for the d$_1$ and d$_2$ directions in Fig.~\ref{fig:Cryst3_4_half_2_d1d2_curves}, exemplified by the saturation field $\mu_0H_S=1.97~\text{T}$ (dashed and solid vertical lines).

\begin{figure}
\centering
\includegraphics[width=\columnwidth]{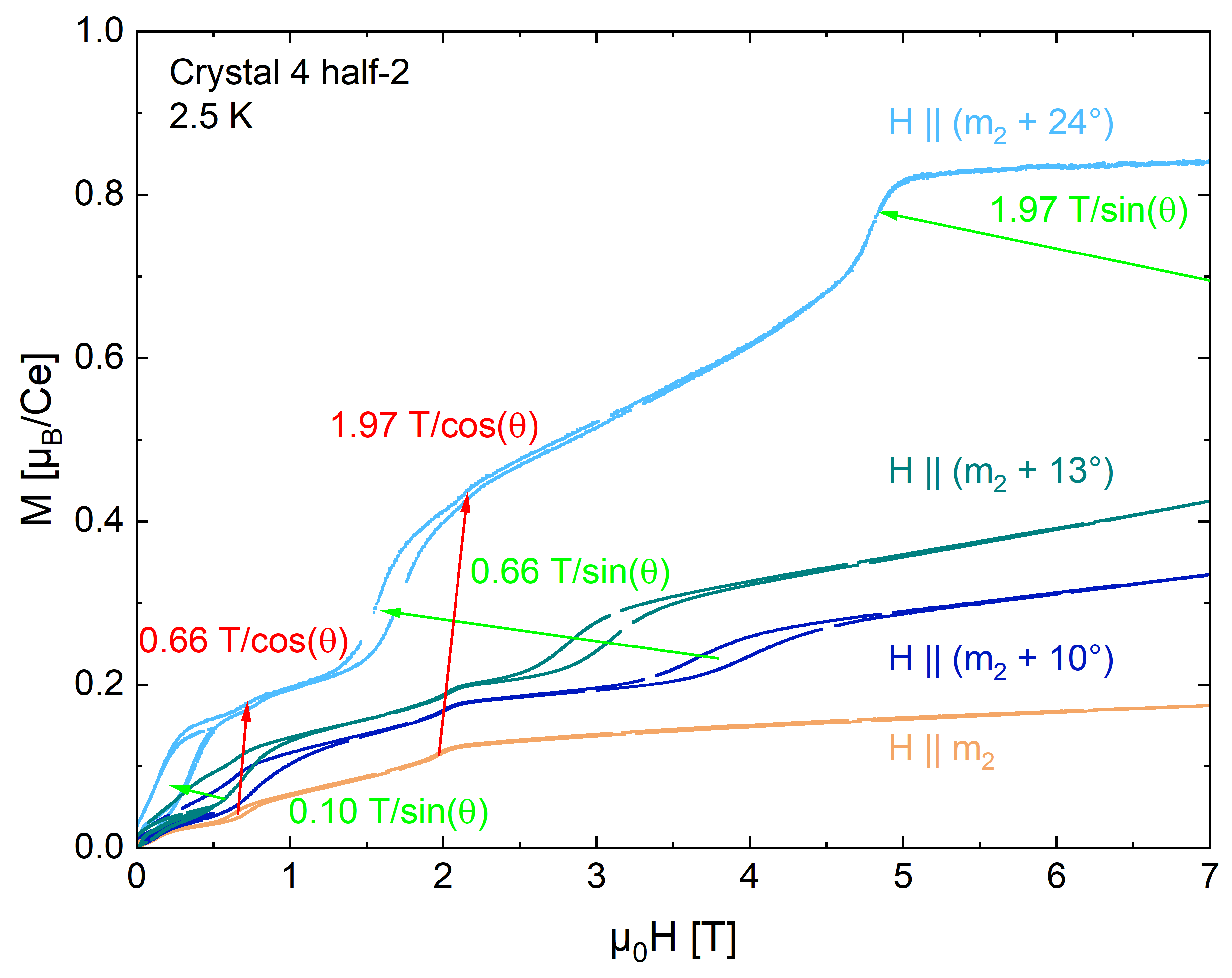}
\caption{Off-axis $M(H)$ curves showing the cosine (red arrows) and sine (green arrows) shift of transition fields for three different rotation angles ($\gamma=10^\circ,\,13^\circ,\,\text{and}\,24^\circ$) with respect to m$_2$.}

\label{fig:transition_shift_angle}
\end{figure}

Off-axis $M(H)$ curves (Fig.~\ref{fig:transition_shift_angle}) reveal that the apparent additional transitions reported in earlier works~\cite{Zhang_CeSb2_magnetization_2017, Liu_CeSb2_neutron_scattering_2020} arise naturally from this cosine/sine shifting of the easy-axis transitions ($0.10~\text{T}$, $0.66~\text{T}$ and $1.97~\text{T}$) with rotation angle. The measurements, performed with clockwise rotation angles starting from $m_2$, follow the predicted cosine/sine dependence. For an equivalent counterclockwise rotation away from $m_2$, identical curves would be expected due to the symmetry of the model. At $m_2 + 45^\circ = m_1 + 45^\circ$, the transitions overlap almost perfectly, as predicted. Our alignment accuracy is within approximately $2^\circ$, which can explain small deviations in the saturation magnetization $M_S$ (as seen for $d_1$ and $d_2$ in Fig.~\ref{fig:Cryst3_4_half_2_d1d2_curves}, top and bottom) and apparent splitting of transition fields (as visible for $d_1$ in Fig.~\ref{fig:Cryst3_4_half_2_d1d2_curves}, bottom).

\subsection{Magnetic phase diagram}
\begin{figure}
\centering
\includegraphics[width=\columnwidth]{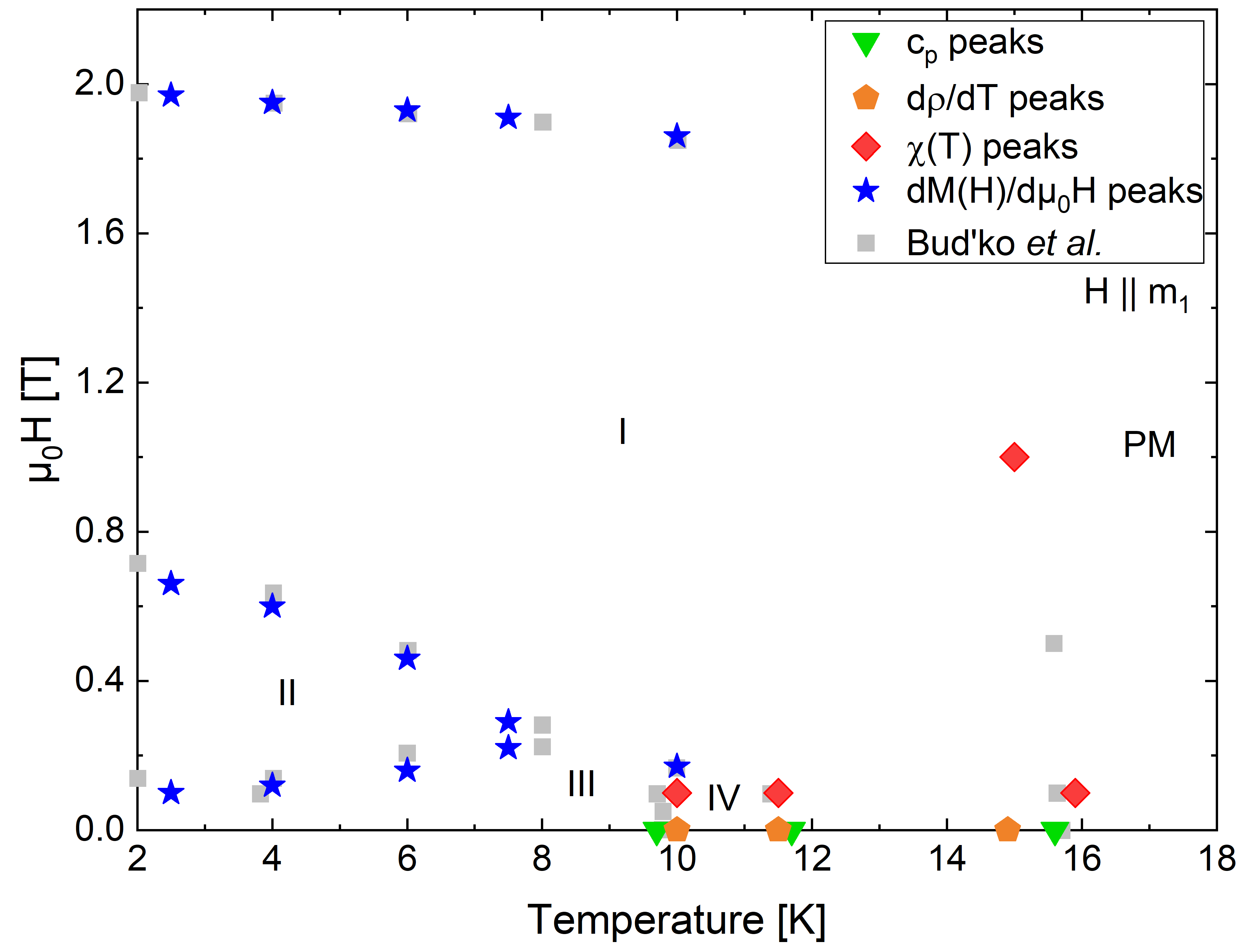}
\caption{Magnetic phase diagram of CeSb$_2$ along the easy axis, constructed from multiple experimental probes and compared with the diagram reported by Bud'ko~\textit{et al.}~\cite{Budko_CeSb2_magnetic_1998}.}
\label{fig:Phasediagram}
\end{figure}

The easy-axis magnetic phase diagram (Fig.~\ref{fig:Phasediagram}) was constructed from heat capacity, resistivity derivative, susceptibility, and magnetization derivative data (Appendix~\ref{Appendix:Magnetic phase diagram data}). Four distinct magnetic phases are identified below 16~K, in agreement with Bud'ko \textit{et al.}~\cite{Budko_CeSb2_magnetic_1998}, as shown by the grey squares. From an analysis of long heat-pulse responses in our heat-capacity measurements, we can exclude that the zero-field transitions are of first order. The susceptibility further suggests an antiferromagnetic nature for phase~IV. Differences from earlier literature phase diagrams can be attributed to misorientation and high twinning (i.e., lower $x$) in previous works \cite{Zhang_CeSb2_magnetization_2017, Liu_CeSb2_neutron_scattering_2020, Trainer_CeSb2_magnetization_2021}.

\subsection{High-temperature phase and twin-reduction with growth technique}
Crossing a structural phase transition during crystal growth can promote twinning, while avoiding such a transition should suppress twinning and improve crystal quality. For CeSb$_2$, a transition from the room-temperature $\alpha$ phase to a structurally unidentified high-temperature $\beta$ phase was proposed in Refs.~\cite{Abulkhaev_CeSb2_Phasediagram_1997, Okamoto_CeSb_Phasediagram_2001} to occur near $(900\pm10)~^\circ$C and to intersect the liquidus line at $\sim93$~at.\%~Sb. 
Independently, pressure-dependent studies have identified a YbSb$_2$-type phase stabilized at high pressure, which may extrapolate to ambient pressure at intermediate temperatures of approximately 50--150$~^\circ$C \cite{Podesta_Poster_pressure_2022, Squire_CeSb2_pressure_2023, Hodgson_PhDThesis_CeSb2_pressure_2023, Squire_PhDThesis_CeSb2_pressure_2024}.

To probe the possible existence of high-temperature phases, we performed high-temperature annealing and rapid quenching experiments on single crystals, cooling them from $1100~^\circ$C to liquid nitrogen temperature. However, PXRD patterns of quenched and as-grown crystals are nearly identical upon visual inspection, showing only the $\alpha$-CeSb$_2$ structure and minor Sb impurities and no evidence for a YbSb$_2$-type phase (Appendix~\ref{Appendix:Powder_XRD}). The quenched sample exhibited a single extra peak not attributable to CeSb$_2$ or Sb, but the overall similarity of the patterns suggests that any high-temperature phase, if present, is either highly unstable at room temperature or very closely related structurally to the $\alpha$ phase. We further performed DTA on CeSb$_2$ single crystals up to $1100~^\circ$C which proved challenging due to significant noise and poor reproducibility, likely caused by Sb evaporation and redeposition during heating (Appendix~\ref{Appendix:DTA}). Within these experimental limitations, no distinct or reproducible thermal signatures attributable to a structural phase transition were detected in the temperature range relevant to either the proposed $\beta$ phase or the YbSb$_2$-type phase.

Since no high-temperature phase could be confirmed structurally, we instead tested whether avoiding the proposed $\alpha$--$\beta$ transition during cooling down the melt improves crystal quality. According to Ref.~\cite{Abulkhaev_CeSb2_Phasediagram_1997}, the transition can be bypassed when the Sb content exceeds 93~\%, consistent with the flux-optimization study by Zhang \textit{et al.}~\cite{Zhang_flux_ratio_2021}, who showed improving crystal quality up to 92~\% Sb.
We therefore grew three sets of crystals:  
(i) crossing the transition (Crystals~1--3),  
(ii) possibly crossing it (Crystal~4), and  
(iii) avoiding it entirely (Crystals~5--7).  
Twinning ratios were determined from $M(2.2~\text{T})$ using Eq.~\eqref{eq:x_ratio_approx} and are summarized in Table~\ref{tab:Twinning_Ratios}.

\begin{table}
\centering
\begin{tabular}{|cc|c|c|c|}
\hline
\multicolumn{2}{|c|}{\textbf{Crossed transition}} & \textbf{Ce\,:\,Sb (Melt)} & \textbf{Crystal} & \textbf{$x$} \\
\hline\hline
\multirow{3}{*}{$\quad\blacksquare$} & \multirow{3}{*}{Yes} & \multirow{3}{*}{10\,:\,90} & 1 & 0.65 \\ 
                   & & & 2 & 0.57 \\ 
                   & & & 3 & 0.59 \\ 
\hline
\multirow{2}{*}{$\quad\tikz \fill(0,0)circle (3.0 pt);$} & \multirow{2}{*}{Possibly}& \multirow{2}{*}{5\,:\,95} & 4 half-1 & 0.87 \\ 
                         & & & 4 half-2 & 0.93 \\ 
\hline
\multirow{3}{*}{$\quad\blacktriangle$} & \multirow{3}{*}{No} & \multirow{3}{*}{3\,:\,97} & 5 & 0.95 \\ 
                  & & & 6 & 0.65 \\ 
                  & & & 7 & 0.91 \\ 
\hline
\end{tabular}
\caption{Twinning ratios $x$ for CeSb$_2$ crystals grown under different conditions. Symbols match those used in Fig.~\ref{fig:M_H_all_curves_and_normalized}: Squares---crystals that crossed the proposed high-temperature transition; circles---crystals that may have crossed it; triangles---crystals that avoided it.}
\label{tab:Twinning_Ratios}
\end{table}

To further refine the twinning analysis and obtain a more accurate picture of the crystal quality, the finite hard-axis contribution can be included (Fig.~\ref{fig:CeSb2_twinning_schematic_real_in_plane}) using
\begin{align}
    &M_{m_1}=x\, M_{\text{easy}}+(1-x)\, M_{\text{hard}},\\
    &M_{m_2}=(1-x)\, M_{\text{easy}}+x\, M_{\text{hard}}.
\end{align}
For crystal~5, inserting $M_{m_1}(2.2~\text{T})=1.721\,\mu_{\text{B}}$/Ce, $M_{m_2}(2.2~\text{T})=0.100\,\mu_{\text{B}}$/Ce, and $M_{\text{hard}}(2.2~\text{T})\lesssim0.062\,\mu_{\text{B}}$/Ce yields $x\lesssim0.98$ and $M_{\text{easy}}(2.2~\text{T})\gtrsim1.759\,\mu_{\text{B}}$/Ce. This refined estimate exceeds the value obtained from the simplified saturation method [Eq.~\eqref{eq:x_ratio_approx}; $x=0.95$], demonstrating that our optimized growth and cleaving procedures achieve even higher twinning ratios than those listed in Table~\ref{tab:Twinning_Ratios}, corresponding to exceptionally high structural quality.

The results from Table~\ref{tab:Twinning_Ratios} indicate that avoiding the proposed high-temperature phase generally yields higher $x$ values, corresponding to reduced twinning. Crystals~5 and~7, both obtained from a single large bulk growth ($>300$~mg), exhibit very high twinning ratios and are likely representative of the intrinsic bulk material. Crystal~6, although grown under the same conditions, crystallized separately and is much smaller, suggesting it may not accurately reflect the overall growth characteristics. Additionally, a slower cooling ramp, as used for crystal~4, appears to improve $x$ as well, indicating that both an Sb-rich flux and controlled cooling are key parameters for minimizing twinning in CeSb$_2$.

\section{Summary}

In this work, we established the intrinsic in-plane magnetic anisotropy of CeSb$_2$ by systematically disentangling the effects of structural twinning. Using a combination of magnetization isotherms, rotational magnetometry, and model analysis, we reconstructed the true easy- and hard-axis responses. The intrinsic hard-axis magnetization is nearly linear, strongly suppressed, and comparable in magnitude to the out-of-plane ($c$ axis) response, while the easy axis saturates at $\approx1.82~\mu_\text{B}$/Ce. These findings resolve long-standing discrepancies in the literature where transition fields and anisotropy appeared sample-dependent, demonstrating that most of the variation can be traced to different twinning ratios and unaccounted in-plane misorientation.  

By combining growth experiments with structural and thermal probes, we further demonstrated that crystal quality is strongly linked to the high-temperature growth conditions. Avoiding the proposed $\alpha$--$\beta$ structural transition during solidification, by increasing the Sb content and by slowing the cooling rate, significantly reduced twinning. PXRD and DTA experiments provided no direct evidence for a distinct $\beta$ phase, suggesting that if it exists, it is either structurally very similar to $\alpha$-CeSb$_2$ or unstable at ambient conditions.  

The resulting twinning-controlled, untwinned or weakly twinned single crystals allowed us to construct a consistent low-temperature magnetic phase diagram, clarifying ambiguities in earlier works. The improved knowledge of the intrinsic anisotropy---in particular, the low-temperature saturation magnetization and Curie-Weiss temperatures---provides a robust starting point for crystal-electric field modeling and constrains possible ground-state scenarios of the Ce$^{3+}$ $J=5/2$ multiplet.

More broadly, our results highlight the critical role of structural twinning in all layered rare-earth diantimonides and establish a reliable route to high-quality CeSb$_2$ single crystals. This paves the way for future investigations into the interplay between anisotropic magnetism, pressure-induced superconductivity, and unconventional electronic correlations in this class of materials.

\begin{acknowledgments}
Work at Ames National Laboratory (J.T.W., S.L.B., and P.C.C.) was supported by the U.S. Department of Energy, Office of Basic Energy Sciences, Division of Materials Sciences and Engineering. Ames National Laboratory is operated for the U.S. Department of Energy by Iowa State University under Contract No.~DE-AC02-07CH11358. 
J.T.W., K.K., and C.K. acknowledge funding from the Deutsche Forschungsgemeinschaft (DFG, German Research Foundation) through the CRC/TRR~288 (422213477, Project No. A03). In addition, J.T.W. acknowledges support from the International Laboratory Visits Scholarship awarded by Goethe University Frankfurt am Main.
\end{acknowledgments}

\section*{Data Availability}
The data that support the findings of this article are openly available \cite{Weber_Gude_2026}.
\clearpage
\renewcommand\thefigure{\thesection.\arabic{figure}}
\setcounter{figure}{0}
\appendix

\section{Twinning and magnetic anisotropy}
\subsection{Magnetic susceptibility}
\label{Appendix:Magnetic susceptibility}
\begin{figure}[H]
\centering 
\includegraphics[width=\columnwidth]{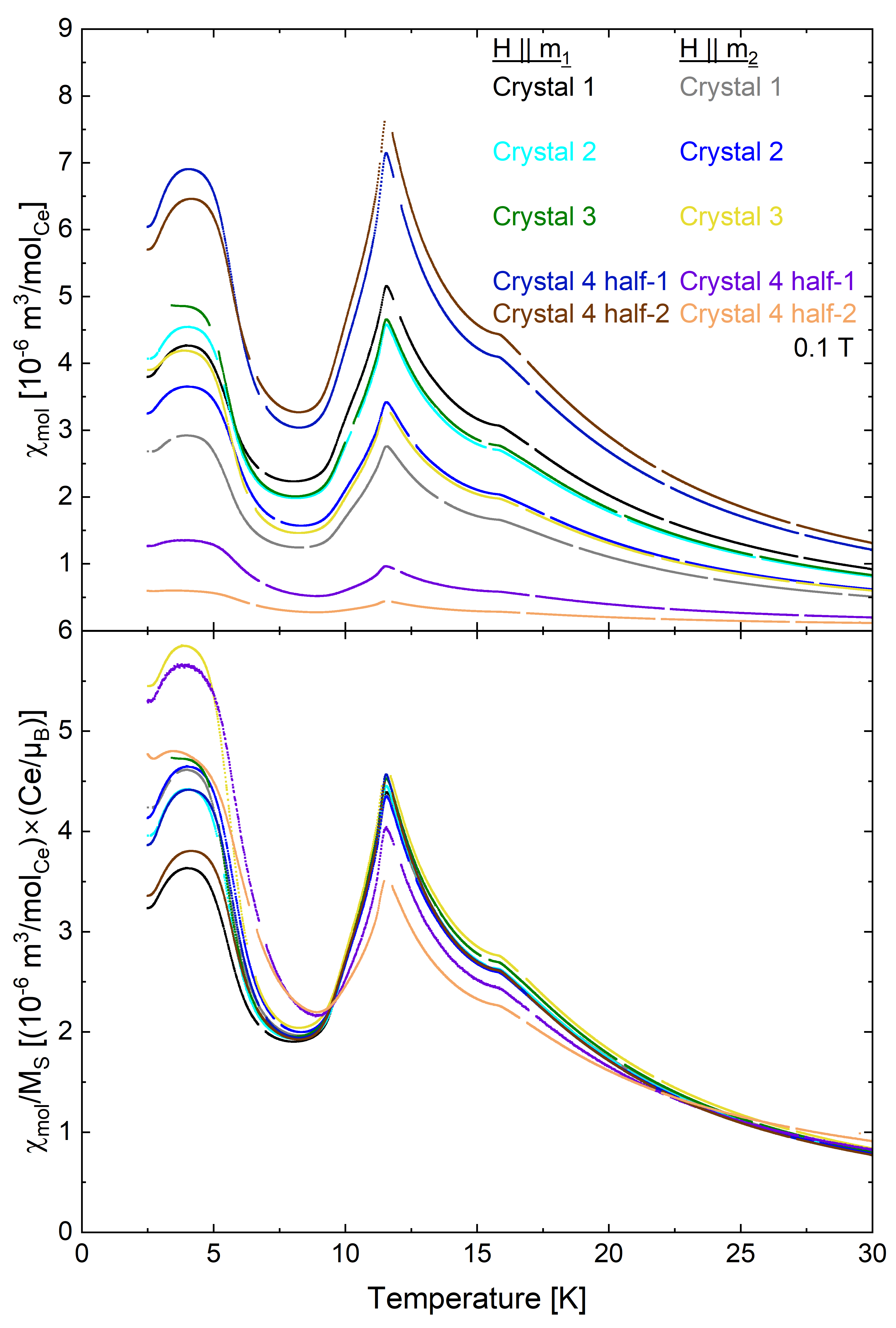}
\caption{Non-normalized (top) and normalized (bottom) molar susceptibility $\chi$ of various crystals measured along m$_1$ and m$_2$ at $\mu_0H=0.1$~T. The data are normalized using the same factors as for the $M(H)$ normalization [$M_S = M(2.2~\text{T})$]. The normalized curves collapse almost perfectly onto one another---except for the hard-axis-dominated ones---demonstrating Ising-type anisotropy, while the twinning picture remains consistent even in the PM regime.}

\end{figure}
\newpage
\subsection{Error analysis of the lower bound of $x$}
\label{Appendix_Error analysis of lower boundary of x}
To estimate the uncertainty in the twinning ratio $x$ derived from Eq.~\eqref{eq:x_ratio_approx}, a Gaussian error propagation was performed. A relative measurement uncertainty of 2~\% was assumed for both magnetization values, and representative values of $M_{m_1}$ and $M_{m_2}$ were used. The resulting propagated relative error, $\Delta x/x$, is given by:
\[\begin{aligned}
x &= \frac{M_{m_1}}{M_{m_1}+M_{m_2}},\\
\Delta
x&=\sqrt{\left(\frac{\partial x}{\partial M_{m_1}}\, \Delta M_{m_1}\right)^2+\left(\frac{\partial x}{\partial M_{m_2}}\, \Delta M_{m_2}\right)^2},\\
\frac{\Delta x}{x}&=\frac{\sqrt{(M_{m_2}\, \Delta M_{m_1})^2+(M_{m_1}\, \Delta M_{m_2})^2}}{M_{m_1}(M_{m_1}+M_{m_2})},\\
\end{aligned}
\]
with
\[
\begin{aligned}
&\frac{\Delta M_{m_1}}{M_{m_1}}=\frac{\Delta M_{m_2}}{M_{m_2}}=2~\%,\\
&1.0\,\mu_{\text{B}}\text{/Ce}\leq M_{m_1} \leq 1.8\,\mu_{\text{B}}\text{/Ce},\\
&0.1\,\mu_{\text{B}}\text{/Ce}\leq M_{m_2} \leq 1.0\,\mu_{\text{B}}\text{/Ce} \\
&\Longrightarrow \frac{\Delta x}{x} \leq 2~\%.
\end{aligned}\]
\subsection{Measurement accuracy}
\label{Appendix:Measurement_accuracy}
\setcounter{table}{2}
\renewcommand{\thetable}{A\arabic{table}}
\begin{table}[H]
\centering
\begin{tabular}{|c|c|c|c|c|}
\hline
\textbf{Crystal} & $M_{m_1}(2.2~\text{T})$ & $M_{m_2}(2.2~\text{T})$ & $M_{m_1}(2.2~\text{T})+M_{m_2}(2.2~\text{T})$ \\
\hline
& \multicolumn{3}{c|}{[$\mu_{\text{B}}$/Ce]}\\
\hline\hline
1 & 1.174 & 0.633 & 1.807 \\ 
\hline
2 & 1.028 & 0.786 & 1.814 \\ 
\hline
4 half-1 & 1.564 & 0.239 & 1.803 \\ 
\hline
4 half-2 & 1.698 & 0.125 & 1.823 \\ 
\hline
5 & 1.721 & 0.100 & 1.821 \\ 
\hline
6 & 1.201 & 0.635 & 1.836 \\ 
\hline
7 & 1.603 & 0.169 & 1.772 \\ 
\hline
\end{tabular}
\caption{Extracted $M(2.2~\text{T})$ values from all measured crystals at 2.5~K. Together with the corresponding sum $M_{m_1}(2.2~\text{T})+M_{m_2}(2.2~\text{T})$. The highest deviation is $(1.836-1.772)/1.772<4~\%$.}
\label{tab:magnetization_sum}
\end{table}
\subsection{Deriving the hard-magnetization response}
\label{Appendix:hard-magnetization}
Formally, $M_{\text{hard}}(H)\, x = M_{m_2}(H) - \alpha M_{m_1}(H)$, with $\alpha$ chosen to suppress metamagnetic steps arising from the easy axis. The intrinsic $M_{\text{hard}}$ is then obtained by dividing by the twinning ratio $x$.
\subsection{Range approximation for the easy- and hard-magnetization axis}
\label{Appendix:upper_lower_bound}

For $M_{\text{easy}}(2.2~\text{T})$, the lower bound is defined by the highest measured raw m$_1$ curve (crystal~5). The upper bound is obtained by dividing the same curve by its twinning ratio $x=0.95$, which itself represents only a lower limit; hence the resulting value is strictly an upper limit.\\

For $M_{\text{hard}}(2.2~\text{T})$, the upper bound corresponds to the largest reasonable difference curve obtained from m$_2$ of crystal~5 after correction with its twinning ratio $x=0.95$
(again a lower bound itself). The lower bound corresponds to the smallest reasonable difference curve obtained without applying the $x$ correction.

\subsection{Inverse magnetic susceptibility}
\label{Appendix:Inverse magnetic susceptibility}
\setcounter{figure}{5}
\begin{figure}[H]
\centering 
\includegraphics[width=\columnwidth]{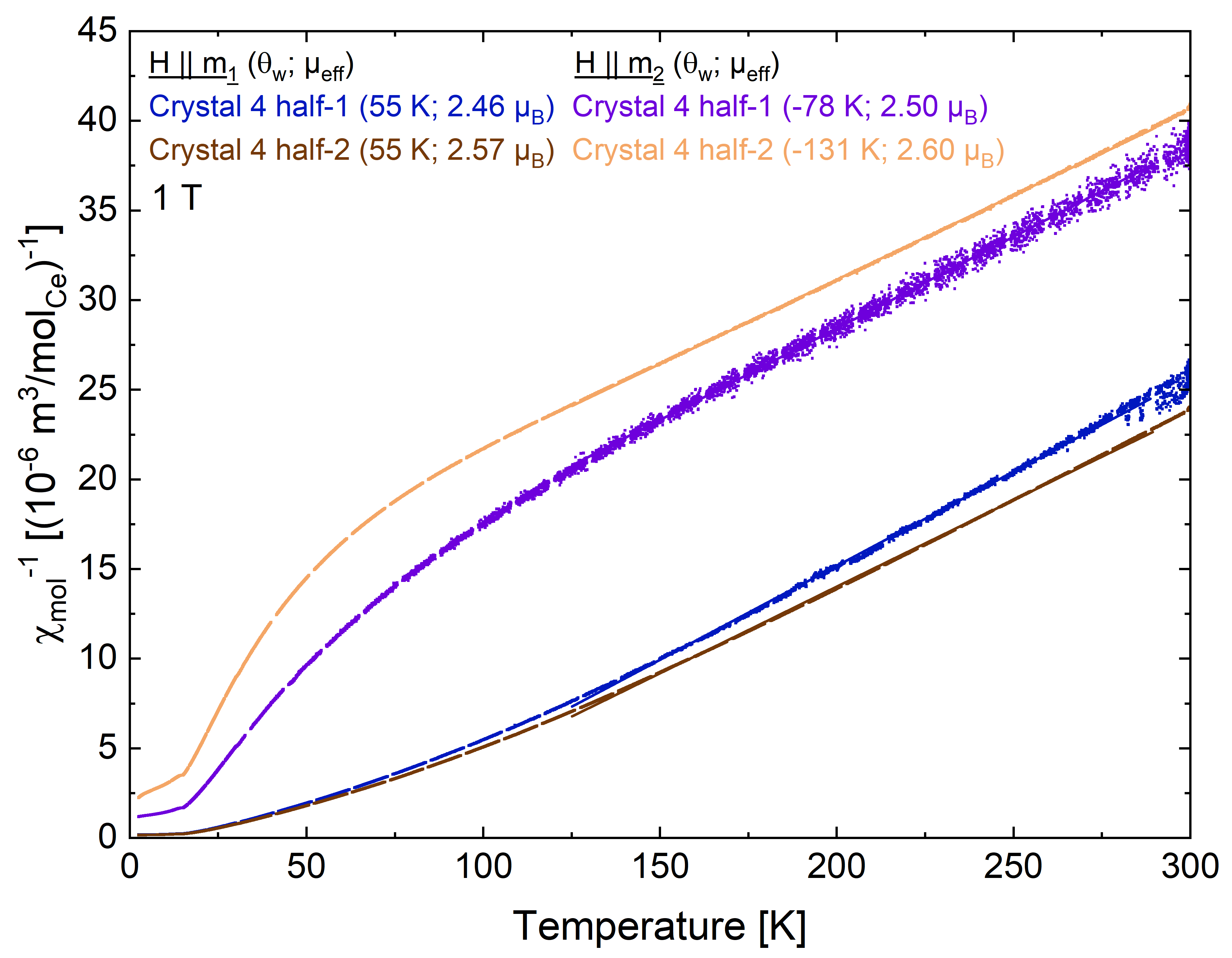} 
\caption{Inverse magnetic susceptibility for different crystals with high $x$, measured along m$_1$ and m$_2$ at $\mu_0H=1$~T. Curie-Weiss fits in the range 125--300~K yield similar effective moments but reveal opposite trends in the Curie-Weiss temperatures: negative for the m$_2$ curves (dominated by the hard axis) and positive for the m$_1$ curves.} 
\end{figure}
\section{Rotational dependence, model validation, and twinning domains}
\subsection{Rotational dependence model}
\label{Appendix:Rotational Dependence Model}
The field-polarized regime is described by
\[
\begin{aligned}
    M_{\text{real,FP}}&=\pm M_{\text{S}},\\
    M_{\text{measured,FP}}&=M_{\text{real,FP}} \, \cos{\theta}\\
                      &= \pm M_{\text{S}} \, \cos{\theta}\\
                      &= M_{\text{S}} \, |\cos{\theta}|,\\
    M_{\text{measured,FP,total}}&=A \, |\cos{\theta}|+B \, |\sin{\theta}|. 
\end{aligned}
\]

The paramagnetic regime is described by
\[
\begin{aligned}
H_{\text{eff}} &= H_{\text{ext}} \, \cos{\theta}, \\
M_{\text{real,PM}} &= \chi_{\text{easy}} \, H_{\text{eff}}, \\
M_{\text{measured,PM}} &= M_{\text{real,FP}} \, \cos{\theta} \\
&= \chi_{\text{easy}} \, H_{\text{ext}} \, \cos^{2}{\theta}, \\
M_{\text{measured,PM,total}} &= A \, \cos^{2}{\theta}+B \, \sin^{2}{\theta}.
\end{aligned}
\]
\section{Magnetic phase diagram}
\label{Appendix:Magnetic phase diagram data}
\setcounter{figure}{0}
\begin{figure}[H]
\centering 
\includegraphics[width=\columnwidth]{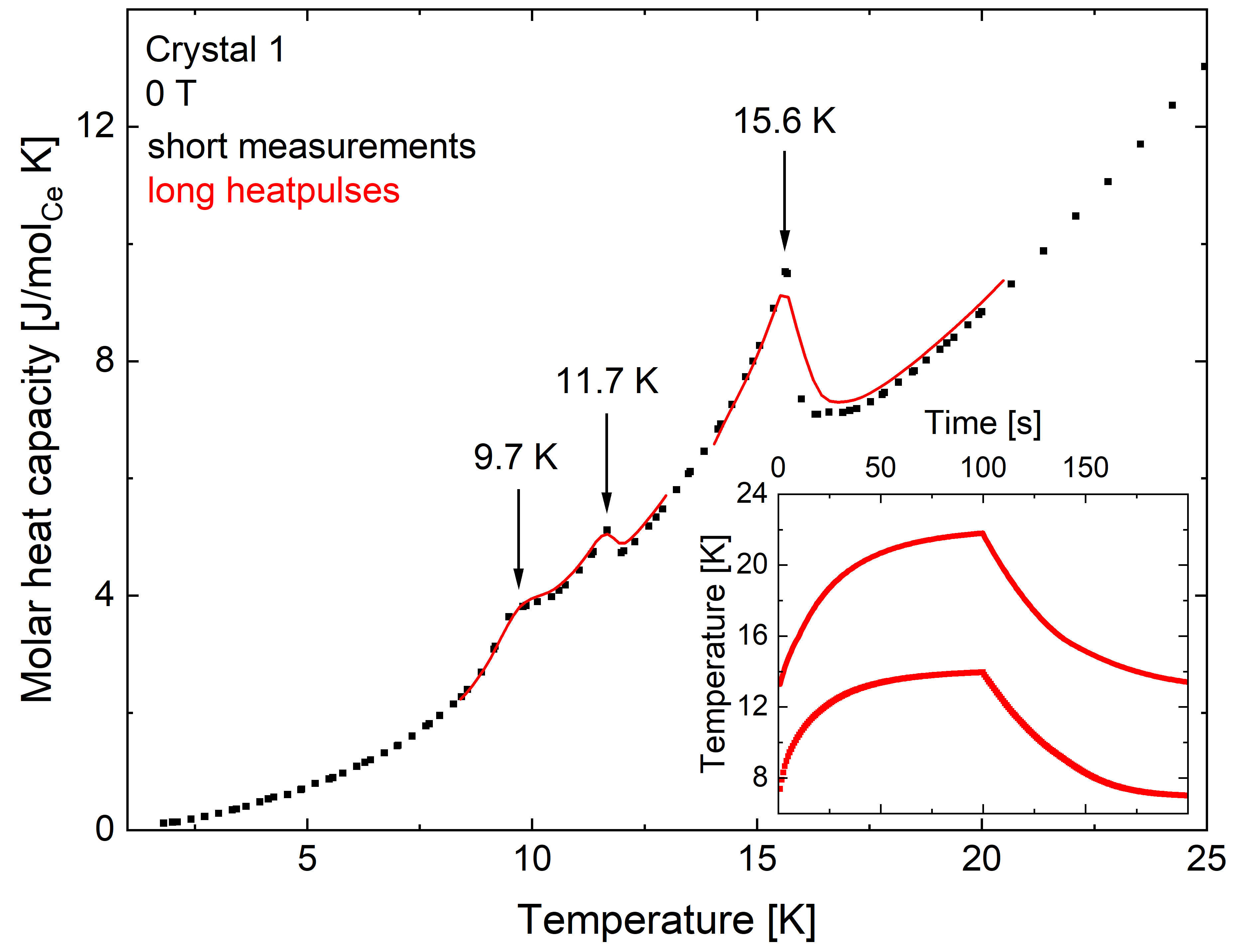} 
\caption{Heat capacity of crystal~1 in zero field measured with short and long heat pulses. Three transitions are observed at 9.7~K, 11.7~K, and 15.6~K. Inset: Long-pulse data show no signatures of first-order behavior at any transition.}
\end{figure}

\begin{figure}[H]
\centering 
\includegraphics[width=\columnwidth]{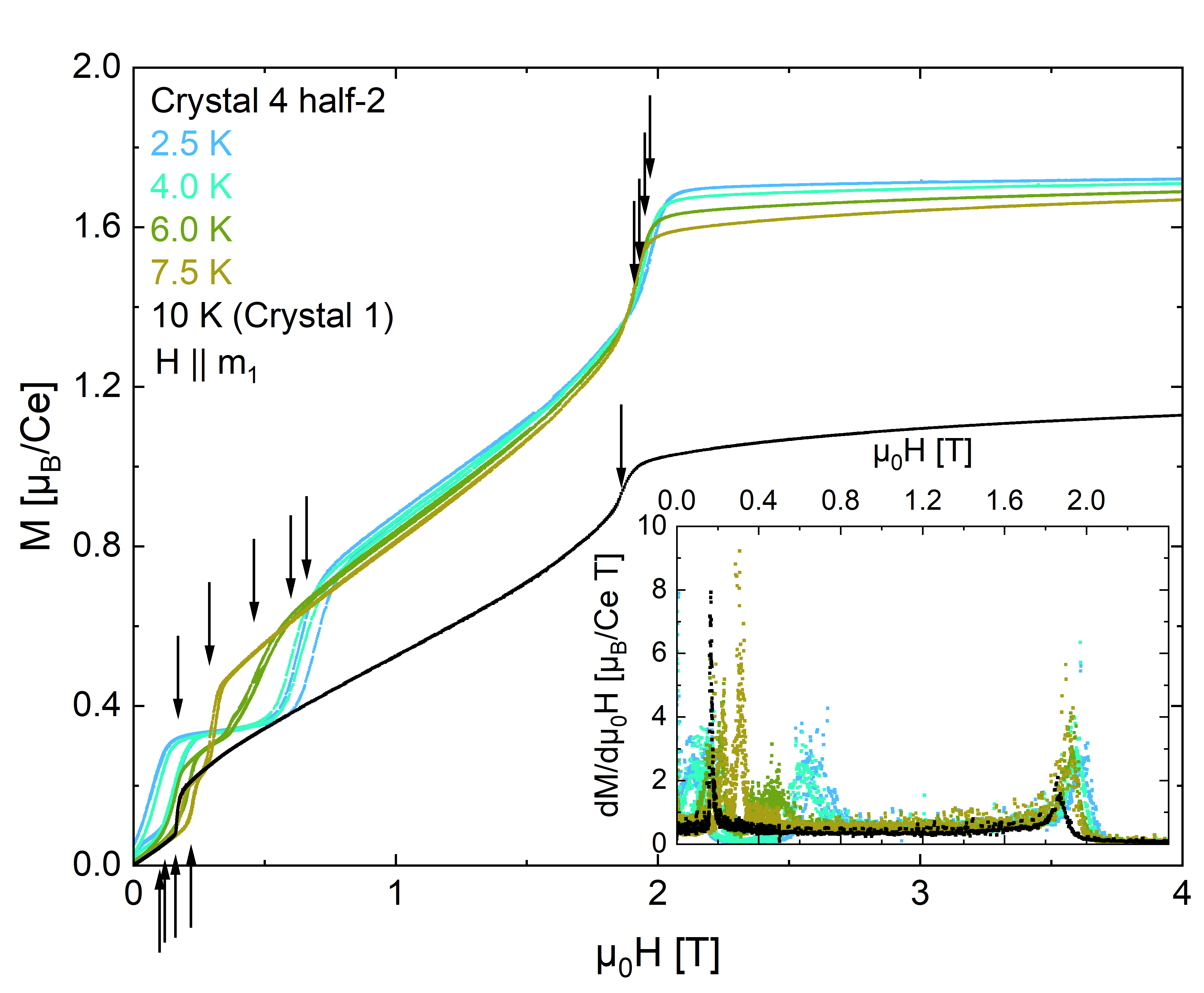} 
\caption{$M(H)$ curves for crystal~4 half-2 (2.5--7.5~K) and crystal~1 (10~K) with $H\parallel$~m$_1$. Transition fields (marked with arrows) were extracted from peaks in $dM/dH$ (inset).}
\end{figure}
\newpage
\begin{figure}[H]
\centering 
\includegraphics[width=\columnwidth]{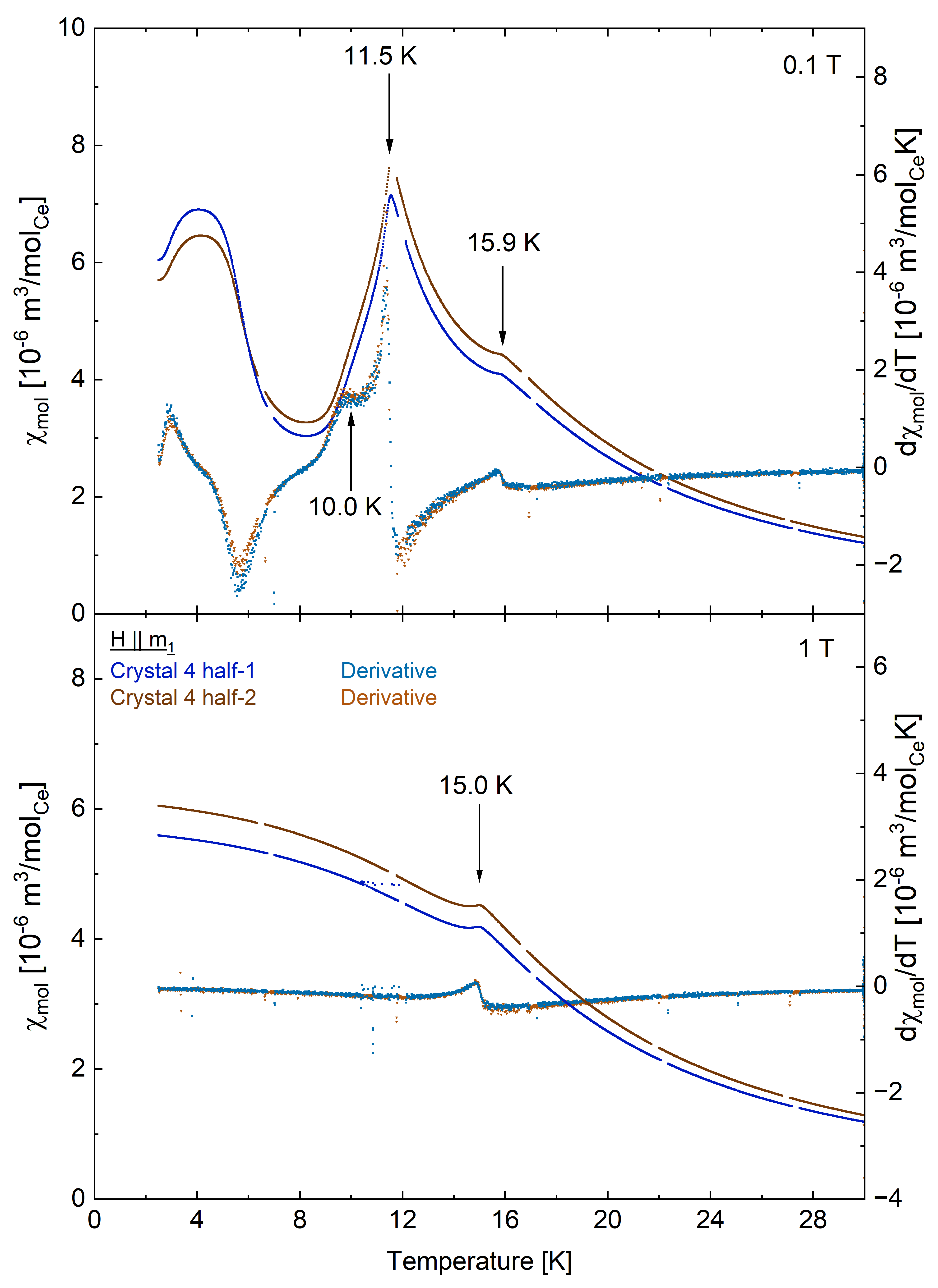}  
\caption{Molar susceptibility of crystal~4 half-1 and half-2 with $H\parallel$~m$_1$ at $\mu_0H=0.1$~T (top) and $\mu_0H=1$~T (bottom). Transitions are marked at 10.0~K, 11.5~K, and 15.9~K for 0.1~T and at 15.0~K for 1~T. The pronounced anomaly at 11.5~K is consistent with an antiferromagnetic transition.}
\end{figure}

\begin{figure}[H]
\centering 
\includegraphics[width=\columnwidth]{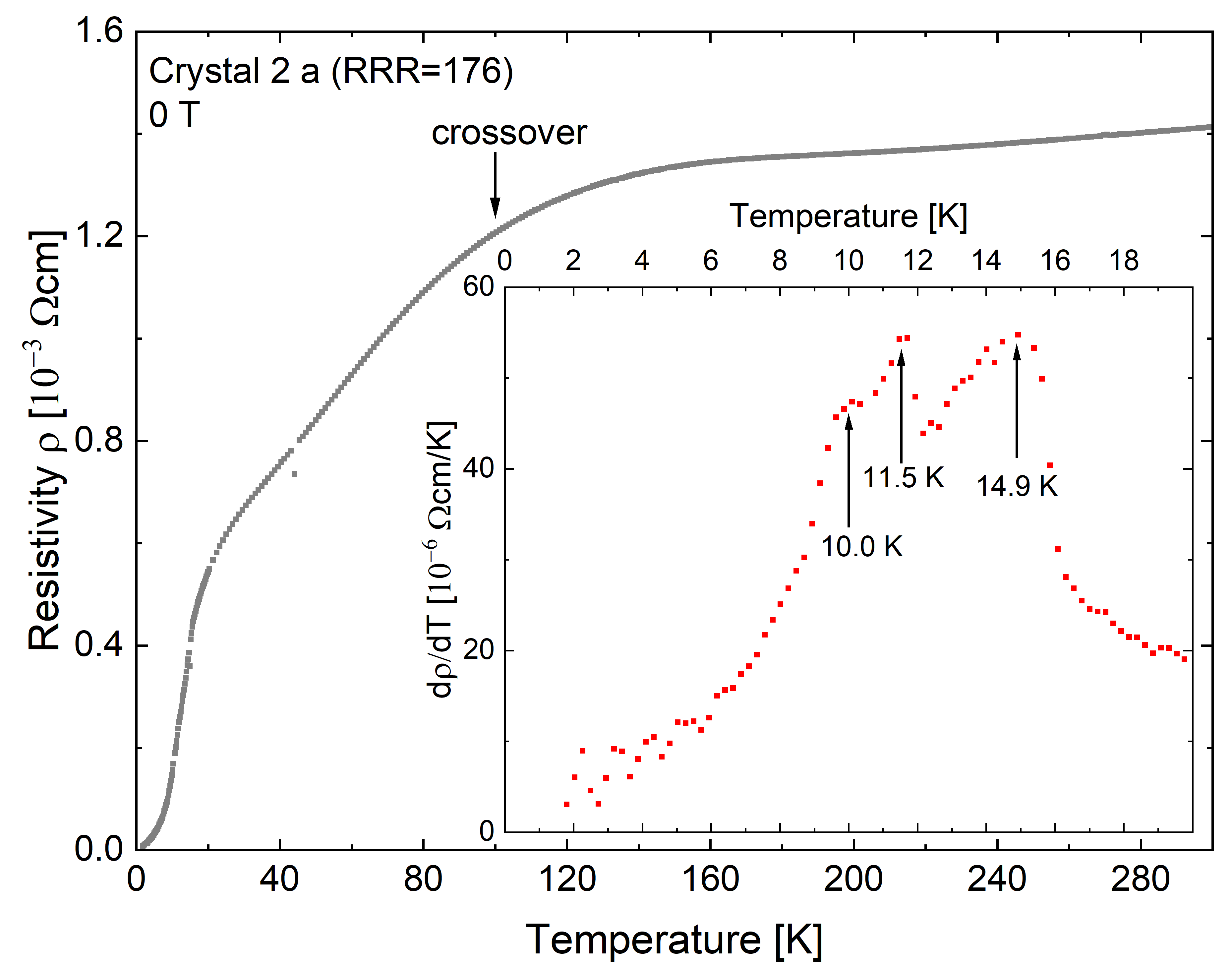} 
\caption{Resistivity of a crystal from the same batch as crystals~1 and 2, showing a crossover near 100~K. Inset: Derivative highlighting transitions at 10.0~K, 11.5~K, and 14.9~K.}
\end{figure}
\newpage
\section{High-temperature phase and twin-reduction with growth technique}
\subsection{Powder XRD}
\label{Appendix:Powder_XRD}
PXRD measurements were performed on as-grown and quenched CeSb$_2$ single crystals from the same batch as crystals~1 and 2 and are compared to simulated patterns for orthorhombic CeSb$_2$ \cite{Wang_CeSb2_structure_1967}, CeSb$_2$ in the YbSb$_2$-type structure \cite{Wang-Steinfink_YbSb2_structure}, and elemental Sb \cite{Barrett_Sb_structure_1962} in Fig.~\ref{Fig:PXRD}. Rietveld refinements proved challenging due to the poor quality of the measurement but are consistent with the presence of only the $\alpha$-CeSb$_2$ phase and elemental Sb. Visual inspection of the diffraction patterns shows that none of the characteristic CeSb$_2$ reflections (notably in the ranges $29.5$–$30^\circ$ and $34.5$–$36^\circ$) weaken or disappear in the quenched sample, as would be expected if a YbSb$_2$-type phase replaced the $\alpha$ phase in a significant amount. In addition, no characteristic peak appears at $30.27^\circ$, and the CeSb$_2$ main peak near $30.7^\circ$ does not shift to a higher angle, as would be expected for the YbSb$_2$-type structure. Any changes in peak intensities are instead attributed to an increased amount of crystalline elemental Sb. The small additional peak near $31^\circ$ in the quenched sample (arrow) cannot be identified and may indicate traces of a high-temperature $\beta$ phase. No other peak shifting, appearance, or disappearance is observed.
\setcounter{figure}{0}
\begin{figure}[H]
\label{Fig:PXRD}
\centering 
\includegraphics[width=\columnwidth]{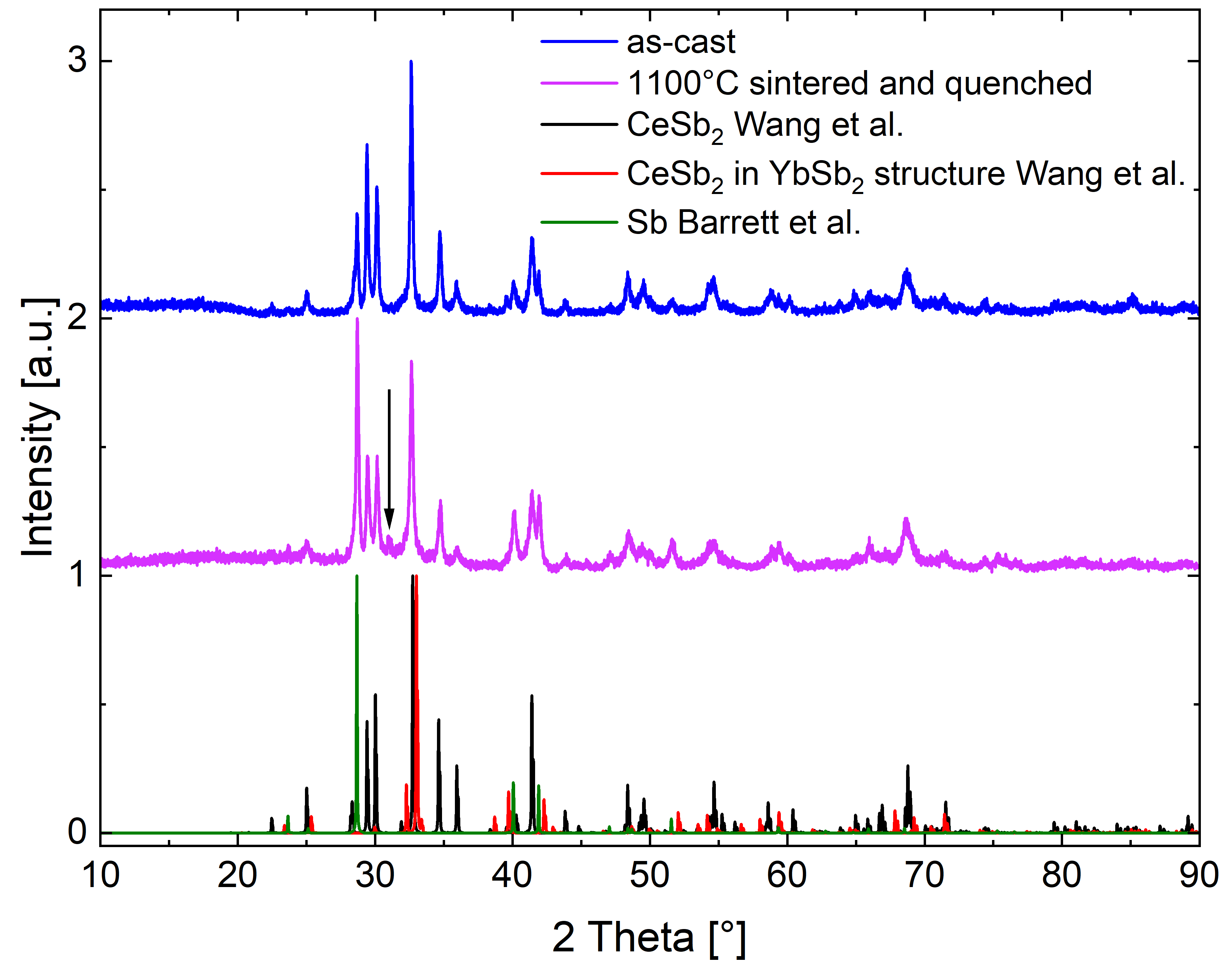} 
\caption{Powder XRD patterns of as-grown and quenched CeSb$_2$ single crystals from the same batch as crystals~1 and 2. Simulated patterns for orthorhombic CeSb$_2$ \cite{Wang_CeSb2_structure_1967}, CeSb$_2$ in the YbSb$_2$-type structure \cite{Wang-Steinfink_YbSb2_structure}, and elemental Sb \cite{Barrett_Sb_structure_1962} are shown for comparison.}
\end{figure}
\subsection{DTA}
\label{Appendix:DTA}
\begin{figure}[H]
\centering 
\includegraphics[width=\columnwidth]{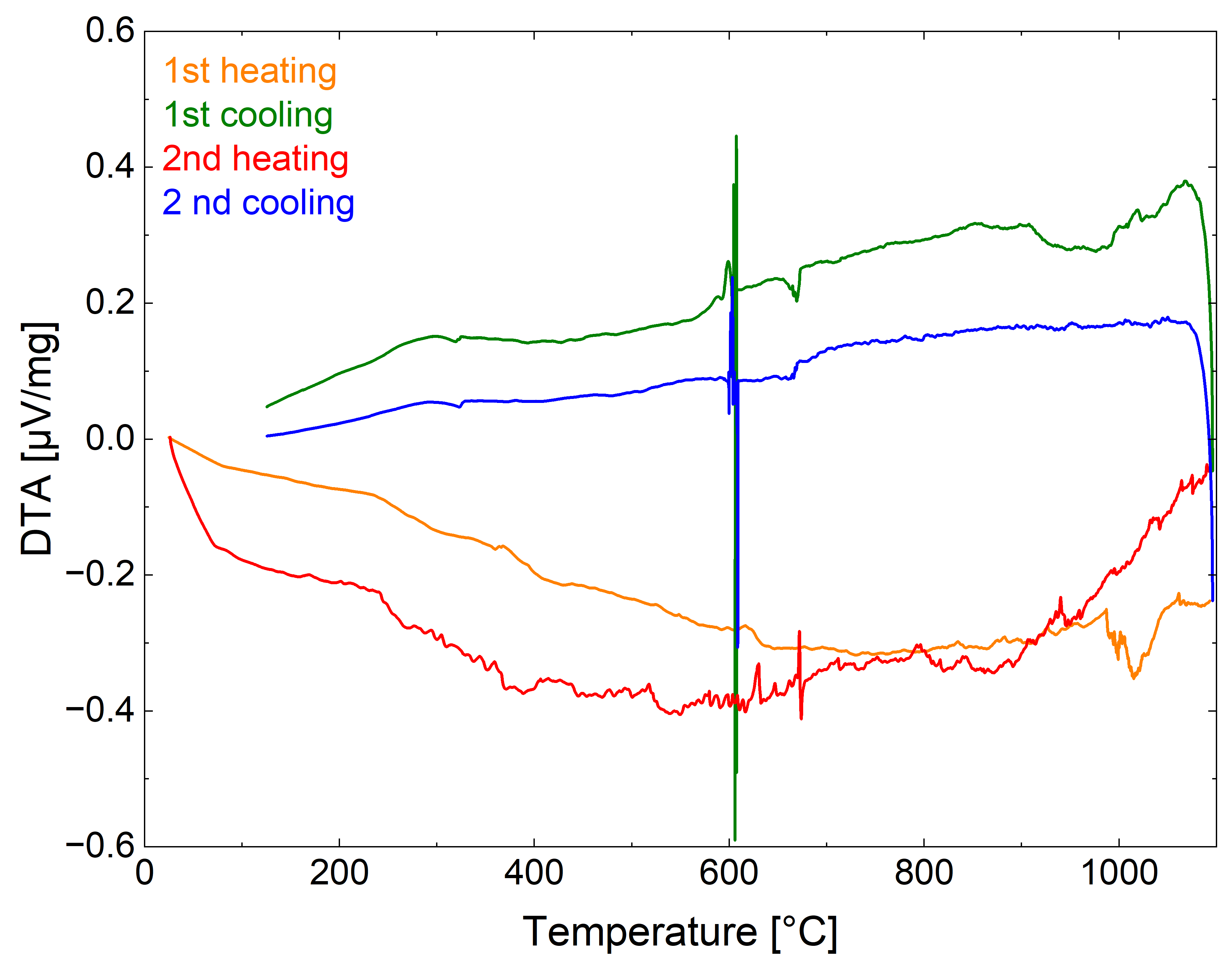} 
\label{Fig:DTA}
\caption{Two consecutive DTA measurements performed on single crystals from the same batch as crystal~3.
The first measurement is shown in orange (heating) and green (cooling), while the second measurement is shown in red (heating) and blue (cooling). The cooling curves are only recorded down to $125~^\circ$C due to passive cooling.}
\end{figure}

Two consecutive DTA measurements were performed on CeSb$_2$ single crystals (Fig.~\ref{Fig:DTA}). A noticeable increase in noise is observed during the second run, indicating progressive degradation of the measurement conditions. No distinct or reproducible thermal signatures attributable to latent heat from a structural phase transition are observed in the temperature ranges relevant to either the proposed high-temperature $\beta$ phase ($\sim900~^\circ$C) or the YbSb$_2$-type phase suggested at intermediate temperatures (50--150$~^\circ$C). The only clear reproducible features appears upon cooling at approximately $T \approx 610~^\circ$C. These features are tentatively attributed to Sb-related evaporation and redeposition. Additional weak features in the cooling curves near $T \approx 330~^\circ$C and $T \approx 665~^\circ$C cannot be assigned at present.
\newpage

\bibliography{CeSb2}

\end{document}